\documentclass[11pt]{article}
\pdfoutput=1
\usepackage{amsmath}
\usepackage{amssymb}
\usepackage{multirow}

\usepackage{graphicx,bbm,mathrsfs}
\usepackage{nicefrac}
\usepackage{bbm}
\usepackage{geometry}       
\usepackage{jheppub}     
\usepackage{bm}        
\usepackage{graphicx,mathrsfs}
\usepackage{nicefrac}
\usepackage{color}
\definecolor{orange}{RGB}{ 148,0,211}

\def\ie{{\it i.e.}}

\newcommand{\be}{\begin{equation}}  
\newcommand{\ee}{\end{equation}}  
\newcommand{\bea}{\begin{eqnarray}}  
\newcommand{\eea}{\end{eqnarray}}  

\newcommand{\tr}{\operatorname{tr}}

\newcommand{\A}{\mathcal A}
\newcommand{\B}{\mathcal B}

\renewcommand{\H}{\mathcal H}
\newcommand{\F}{\mathcal F}

\newcommand{\Tr}{\operatorname{Tr}}

\newcommand{\bmat}{\begin{pmatrix}}
\newcommand{\emat}{\end{pmatrix}}

\newcommand*{\Scale}[2][4]{\scalebox{#1}{$#2$}}%

\newcommand{\nn}{\nonumber}

\renewcommand{\H}{\mathcal{H}}

\begin{document}
                
\title{The Excitation of the Global Symmetry-Breaking Vacuum in Composite Higgs Models}

\author{Sylvain~Fichet$\,^a$,}
\emailAdd{sylvain.fichet@gmail.com}

\author{Gero von Gersdorff$\, ^b$,}
\emailAdd{gersdorff@gmail.com}

\author{Eduardo Pont\'on$\,^a$ and}
\emailAdd{eponton@ift.unesp.br}

\author{Rogerio Rosenfeld$\,^a$}
\emailAdd{rosenfel@ift.unesp.br}

\affiliation{$^a$ ICTP South American Institute for Fundamental
  Research \& Instituto de F\'isica Te\'orica \\ 
  Universidade Estadual  Paulista, S\~ao Paulo, Brazil}

\affiliation{$^b$ Departamento de F\'isica, Pontif\'icia Universidade
Cat\'olica de Rio de Janeiro, Rio de Janeiro, Brazil}

\abstract{ We consider scenarios of Higgs compositeness where the
Higgs doublet arises as a pseudo-Nambu Goldstone boson.  Our focus is
the physical scalar (``radial") excitation associated with the global
symmetry breaking vacuum, which we call the \textit{global Higgs}.
For the minimal case of a $SO(5)/SO(4)$ coset, the couplings of the
global Higgs to Standard Model (SM) particles are fully determined by
group theoretical factors and two decay constants.  The global Higgs
also couples to the composite resonances of the theory, inducing an
interaction with the SM gauge bosons at one-loop.  We thoroughly
analyze representative fermionic sectors, considering a global Higgs
both in the $\bf 5$ and $\bf 14$ representations of $SO(5)$ and taking
into account the renormalization group evolution of couplings in the
composite sector.  We derive the one-loop effective couplings and all
decays of the global Higgs, showing that its decay width over mass can
range from ${\cal O} (10^{-3})$ to ${\cal O}(1)$.  Because of the
multiplicity of the resonances, the coupling of the global Higgs to
gluons is sizeable, potentially opening a new window into composite
models at the LHC. }

{\center \today}

\maketitle

\thispagestyle{empty}
\setcounter{page}{1}

\section{Introduction}
\label{se:intro}

It is by now established with a high degree of significance that the
125 GeV resonance discovered at the LHC in July 2012 is an excitation
of the broken electroweak vacuum.  This discovery sheds light on the
nature of electroweak symmetry breaking (EWSB), and is consistent with
an elementary Higgs doublet as posed in the Standard Model of Particle
Physics (SM).

Now that it is clear that a scalar field is present at the TeV scale,
it becomes more urgent to understand why the scale of electroweak
breaking is so small compared to ultra-violet (UV) scales such as the
Planck mass, $M_{\rm Pl}$.  A first approach to tackle this hierarchy
problem is to assume that new particles appear near the electroweak
scale in order to cancel the huge $O(M_{\rm Pl}^2)$ corrections that
tend to destabilize the weak scale.  A second possibility is to assume
that this scalar field exists only at low energies, as a bound state
of more fundamental constituents.  At energies higher than the
compositeness scale $\Lambda$, the EW scalar would dissolve into
constituents of spin higher than zero, which can be immune to large
quadratic corrections.

This second possibility, that interprets the Higgs boson as a
composite object, is at the center of our attention in this work.  The
fact that we have tested the Higgs boson with energies comparable to
its mass without revealing any obvious substructure, so that the
binding energy must be comparable to its rest mass, suggests that the
bound state arises from some underlying strong dynamics, as opposed to
being a weakly bound state.  The generic scenario of strong dynamics
leading to a composite scalar field is, however, in tension with the
observed properties of the Higgs boson.  Indeed, one would generically
expect that such a composite field would feature a broad decay width,
and be accompanied by other, nearby resonances.  However, the Higgs
boson is observed to be narrow, and no other light states
(e.g.~vectorlike fermions or additional spin-1 fields) have been
observed up to now at the LHC.

A class of scenarios that naturally reconciles strong coupling
dynamics with the observed Higgs boson is the one postulating that the
EW scalar doublet is actually the pseudo-Nambu-Goldstone boson (pNGB)
of a global symmetry $G$ spontaneously broken to a subgroup $G'$ at an
energy scale $\hat f$.  A small explicit breaking of $G$ would then be
at the origin of the EW scalar potential, perhaps allowing for a
dynamical understanding of EWSB itself.  In this framework, a mass gap
automatically splits the EW scalar from heavier resonances, and its
couplings are naturally weak so that the Higgs boson has a narrow
width.  These scenarios are referred to as ``pNGB composite Higgs''
models (for a recent review, see \cite{Panico:2015jxa}).  The
magnitude of $\hat f$ is bounded from below by LHC searches for e.g
vectorlike fermions, as well as from deviation from the expected SM
Higgs couplings.  The magnitude of the compositeness scale $\Lambda$
can also be bounded from below using Higgs coupling measurements.

Although composite Higgs models usually assume an underlying
strongly-coupled dynamics, their UV completion has received relatively
little attention.  Instead, most of the literature focuses on the
low-energy effective theory describing the Higgs boson properties
below the compositeness scale.  More precisely, the standard way to
proceed is to work within the non-linear $\sigma$-model of the $G/G'$
coset.  This effective description is fully appropriate at low
energies, when the $G'$ vacuum remains unperturbed.  In contrast,
whenever the $G'$ vacuum can be excited, the corresponding degrees of
freedom must be included in the $\sigma$-model.  This implies that a
particle with the quantum numbers of the vacuum, \ie~a new neutral,
CP-even scalar is potentially present in the effective field theory.
We shall refer to this scalar as the \textit{global Higgs} throughout
the rest of this paper, and denote it by $\phi$.  The mass $m_\phi$ of
the global Higgs should be smaller than the cutoff of the effective
theory, but apart from that it is a free parameter.

In this paper, we will investigate the conditions under which the
global Higgs can arise and what are its properties.  The possible
presence of a global Higgs in the composite Higgs framework seems
rather intriguing, and to the best of our knowledge, has so far only
been studied in Refs.~\cite{Buttazzo:2015bka,Feruglio:2016zvt}.  We will go beyond the
simple models studied in  \cite{Buttazzo:2015bka, Feruglio:2016zvt} by including couplings
to spin-1 resonances, and also study more general fermion sectors as
well as the possibility of embedding the global Higgs into non-minimal
$SO(5)$ representations.  Moreover, we formulate the theory entirely
in terms of the nonlinear variables, allowing for a more direct
comparison to the usual literature on composite Higgs models.

In Section~\ref{se:scenarios}, we establish a broad picture of the
global Higgs properties based on general arguments.  Focusing on the
$SO(5)/SO(4)$ coset, we derive the bosonic couplings of the global
Higgs in Sec.~\ref{se:boson_couplings}.  We then define a set of
benchmark scenarios for the fermionic sector in
Sec.~\ref{se:fermion_couplings}, and compute systematically the 1-loop
effective couplings of the global Higgs to SM gauge bosons in
Sec.~\ref{se:oneloop}.  The renormalized couplings of the composite
sector are computed in Sec.~\ref{se:perturbativity}, and the decay
widths and branching fractions are presented in Sec.~\ref{se:decays}.
Section~\ref{se:conclusions} contains our conclusions.

\section{A Global Higgs-like Scalar in the Composite Higgs Paradigm}
\label{se:scenarios}

Composite Higgs models -- where the EW symmetry is broken by the
condensation of pseudo-NGB's arising from the spontaneous breaking of
an approximate global symmetry at a higher scale -- are typically
studied in the low energy regime, below the scale of global symmetry
breaking.  As such, one needs only to parametrize the pNGB degrees of
freedom, thereby implementing the global symmetry non-linearly.
Presumably, such a low-energy description is obtained by integrating
out heavy modes.  These heavy states include the global Higgs, scalar
radial modes that, together with the pNGB's, would enter in a complete
$G$ multiplet (denoted by $\Phi$) and would allow a linear
implementation of the full global symmetry (not just of the unbroken
subgroup $G'$).

An interesting example is provided by the model considered in
Ref.~\cite{vonGersdorff:2015fta}.  In that case, the breaking of the
global symmetry is induced by 4-fermion interactions with a
coefficient near criticality.  Indeed, allowing for a mild tuning of
this coefficient so that it lies slightly above a certain critical
value, the symmetry breaking mechanism can be equivalently described
by the condensation of a (composite) scalar in a complete
$G$-representation,\footnote{In the model studied
in~\cite{vonGersdorff:2015fta}, the $G$-symmetry was $SO(5)$, and the
scalar was in the fundamental of $SO(5)$.  After condensation, the
symmetry is broken to $G' = SO(4)$, generating 4 (p)NGB's plus one
real scalar, the global Higgs.  In other non-minimal examples one can
have both additional pNGB's, as well as additional massive scalar
degrees of freedom.  One such example is the breaking of $SO(5)$ by
the {\bf 14} representation, which in addition to the massive $SO(4)$
singlet ``radial" mode, has an additional massive symmetric tensor of
$SO(4)$ in its spectrum.} such that there is a hierarchy between the
symmetry breaking scale, $\hat{f}$, and the cutoff $\Lambda$
associated with the non-renormalizable 4-fermion interactions.  The
global Higgs typically has a mass $m_{\phi}$ of the order of the
symmetry breaking scale.  Interestingly, some fermion resonances are
expected to have masses of order $m_{\phi}$ or somewhat
below.\footnote{As in models of top
condensation~\cite{Terazawa:1976xx,Bardeen:1989ds}, there is a definite relation
between the global Higgs mass and the dynamical mass of the fermions
that bind together to form the global Higgs.} On the other hand, as
explained in~\cite{vonGersdorff:2015fta}, spin-1 resonances associated
with the underlying strong dynamics are expected to be heavier, with
masses of order $\Lambda$.

Thus, one can be in a situation where some of the fermionic
resonances, in addition to the global Higgs, may be more readily
accessible than other higher spin excitations.  The collider
phenomenology of such fermion states has been widely studied in the
context of general composite Higgs
scenarios~\cite{Contino:2006qr,Contino:2008hi,AguilarSaavedra:2009es,Mrazek:2009yu,DeSimone:2012fs,Carena:2014ria,Matsedonskyi:2015dns}
and
beyond~\cite{Berger:2012ec,Okada:2012gy,Cacciapaglia:2012dd,Kearney:2013cca,Angelescu:2015kga}.
Here our focus is rather on the properties of the global Higgs.

We start by establishing a picture of the global Higgs properties in
general terms, leaving a more concrete presentation to the following
section.  First of all, since the global Higgs is by definition an
excitation of the $G/G'$ vacuum, it interacts with the pNGBs that
parameterize this vacuum.  Due to the approximate shift symmetry, such
interactions involve covariant derivatives, and one expects tree-level
couplings of the global Higgs to the pNGB's, \ie~to the SM Higgs boson
and to the longitudinal components of the electroweak gauge bosons.
The global Higgs could also in principle couple to the vector
resonances of the strongly interacting sector.  These couplings
introduce another scale in the model, which we call $f_\rho$, and will
be discussed in the following section.

Importantly the global Higgs has couplings to the fermions in the
spectrum and also, via loop effects, to pairs of gluons and photons.
In order to discuss such effects it will be useful to summarize first
the framework of partial compositeness~\cite{Kaplan:1991dc}, which
allows to elegantly accommodate the SM flavor structure within the
composite Higgs paradigm.  One assumes here the presence of an
elementary sector, in addition to the composite sector giving rise to
the Higgs and other resonances.  The elementary sector contains three
families of chiral fermions $q$, $u$, $d$, $l$, $e$,
\footnote{Generation indices are not shown.} and mimics exactly, in
its $SU(2)_L \times U(1)_Y$ quantum numbers, the fermion field content
of the SM. The composite sector, on the other hand, consists of
vector-like states in complete $G$ representations.  Each $SU(2)_L$
multiplet in the elementary sector is associated with a composite
$G$-multiplet $Q$, $U$, $D$, $L$, $E$, which itself contains some
states with the corresponding $SU(2)_L \times U(1)_Y$ quantum numbers.
This allows bilinear mixing between the elementary and composite
sectors, thereby breaking explicitly the global symmetry
$G$.\footnote{Another source of explicit $G$-breaking is the gauging
of $SU(2)_L \times U(1)_Y$ itself.} In this framework, the light mass
eigenstates are identified with the SM fermions.  They are accompanied
by heavy vectorlike ``partners''.
  
The vectorlike masses of the composite fermion sector will be denoted
by $M_{Q}$, $M_{U}$, $M_{D}$, $M_{L}$ and $M_{E}$.  The
composite fermions also have interactions with the pNGB's, which will
eventually give rise to the SM Yukawa couplings.  One therefore often
refers to the Yukawa interactions between the composite states as
``proto-Yukawa'' interactions.  Having embedded the pNGBs into the
$G$-multiplet $\Phi$, proto-Yukawa interactions take the schematic
form \footnote{Note that we are being rather schematic since the
precise contractions between the various fields depend on the
$G$-representations they belong to.  For our present purposes this
will be sufficient (specific examples will be shown in the following
section).}
\bea
{\cal L}_{\rm proto-Y} &=& - \xi_{U} {\cal O}_U(\Phi) \bar{Q} U - \xi_{D} {\cal O}_D(\Phi) \bar{Q} D - \xi_{E} {\cal O}_E(\Phi) \bar{L} E + {\rm h.c.}~,
\label{proto}
\eea 
where the ${\cal O}_X(\Phi)$ are appropriate functions of $\Phi$ such
that the above terms are $G$ invariant.  The elementary-composite
mixing terms take the form
\bea
{\cal L}_{\rm mix} &=& - \Delta_{q} \bar{Q} \cdot q - \Delta_{u} \bar{U} \cdot u - \Delta_{d} \bar{D} \cdot d - \Delta_{l} \bar{L} \cdot l - \Delta_{e} \bar{E} \cdot e + {\rm h.c.}~,
\label{mix}
\eea
where the dot denotes an appropriate projection of the states in the
$G$-multiplets with the correct SM quantum numbers.  The physical SM
states are linear combinations of the elementary and composite states
with mixing angles controlled by the mixing masses $\Delta_{q},
\ldots, \Delta_{e}$.  Only through the bilinear mixing above do the
lightest mass eigenstates acquire interactions with the pNGB's
contained in $\Phi$, thus leading to Yukawa terms as in the
SM.\footnote{The vectorlike masses, Yukawa couplings and mixing
parameters need not be simultaneously diagonal in flavor space but,
for simplicity, this is not reflected in our notation above.}

A scenario that has received much attention -- commonly known as
\textit{anarchy} 
-- assumes that the proto-Yukawa couplings $\xi_{U,D,E} $ are all of
the same order (and of order one to a few).  The observed hierarchies
in the SM fermion spectrum then arise from hierarchies in the mixing
angles above: the lightest fermions are mostly elementary and hence
weakly coupled to the SM Higgs field (the pNGB states in $\Phi$),
while the heavy top has a sizeable component in the composite sector
(\ie~the mixing angle is large).  A second, perhaps less
well-motivated, possibility is that the mixing angles are of the same
order, and instead the SM fermion mass hierarchies arise from
hierarchies in the proto-Yukawa couplings themselves.  We will
consider both possibilities.

Let us now turn to the couplings of the global Higgs to the fermion
sector.  First, since the global Higgs is contained in $\Phi$, it
couples to composite fermion pairs as dictated by the proto-Yukawa
structures of Eq.~\eqref{proto}.  We can thus expect a coupling of the
global Higgs to the heavy mass eigenstates, controlled by the $SO(5)$
Yukawa couplings $\xi_{U,D,E}$.  The global Higgs can also couple to a
SM fermion and one of its vectorlike partners.  Such couplings require
the mixing terms of Eq.~(\ref{mix}).  However, the proto-Yukawa
interactions induce couplings between the global Higgs and SM fermion
pairs only after EWSB. The reason is that the global Higgs is, by
definition, a SM singlet and there are no fermion bilinear singlets in
the SM. The induced couplings will be seen to be proportional to
the SM fermion mass.

As already mentioned, there are loop-level induced interactions
between the global Higgs and gluons or photons.  Although suppressed,
these can play a central role in the global Higgs phenomenology.
These couplings are induced in complete analogy to the SM case,
through loops of colored or charged states whose masses get a
contribution from the breaking of the global symmetry.  The importance
of such effects depends on the size of the proto-Yukawa couplings and
therefore on whether we assume an anarchic scenario or
not.\footnote{Note that, since Eq.~(\ref{proto}) is $G$-symmetric, no
such couplings between gluons/photons and the SM Higgs are induced at
this point, since here the SM Higgs is an exact NGB. Only the global
Higgs enters into the above discussion.  When the mixing angles of
Eq.~(\ref{mix}) are taken into account, couplings between the SM Higgs
and gluons/photons are induced.  The contribution from the elementary
sector is dominated by the top quark, as in the SM. The contributions
from the composite sector, on the other hand, are suppressed by their
large vector-like masses.  The deviations from the SM couplings in
such scenarios have been studied elsewhere (see,
e.g.~\cite{Carena:2014ria}) and are not the focus of our study.} The
point is that the composite fermion masses can receive both symmetry
breaking ($\sim \xi \hat{f}$) and symmetry preserving contributions.
As is well known, whenever the vector-like mass is small compared to
the symmetry breaking contribution, the loop-induced coupling of the
global Higgs to two gauge bosons (through a triangle diagram) exhibits
a non-decoupling behavior that is already apparent when the Yukawa
coupling $\xi$ is of order one.  This behavior remains qualitatively
true when the vector-like mass is comparable to the symmetry breaking
one.  Therefore, whenever the vector-like masses $M_{Q}, \ldots,
M_{E}$ are not much larger than $\xi \hat{f}$, each resonance gives a
comparable contribution, and the net effect can be encoded into an
effective multiplicity factor.  In the first scenario discussed above,
with order one to a few proto-Yukawa couplings and hierarchical mixing
angles, one can therefore expect that the heavy sector associated with
each SM state can give a sizeable effect.  The details depend on the
$G$-representations of the spin-1/2 resonances, which may be different
for the up-quark sector, the down-quark sector and the leptons.  Also,
if one insists to remain in the perturbative regime at the scale of
the global Higgs mass, a large number of resonances can put an upper
bound on the Yukawa couplings $\xi$, so that the symmetry-breaking
effects cannot be arbitrarily large compared to the vectorlike
effects.  We defer further details to the next section.  However, to
get an idea of the size of the multiplicity factors involved, one can
consider what would have been the situation if the SM Yukawa couplings
were all of order one.  In that case, for the gluon fusion process,
for instance, one would have obtained an amplitude about 6 times
larger than the top contribution, a factor that gets squared in the
cross section.  The multiplicity factors in the global Higgs case can
potentially be even larger since the global symmetry structure often
suggests the presence of relatively large representations, as we will
see in the next section.  It is therefore important to analyze such
enhancements in more detail.

On the opposite extreme, \ie~the case where the mixing angles are all
of order one but the proto-Yukawa couplings are hierarchical, one
expects that only the resonances associated with the top sector will
be important.  The resonances associated with the lighter fermions
will give contributions to the loop processes that are suppressed,
much as those of the light fermions in the SM Higgs case.  Hence, this
limit provides a ``minimum" contribution to the 1-loop amplitudes, and
thus we will present it as one of the benchmarks in our study,
regardless of how likely it is to be realized in nature.

In summary, the picture that emerges is that the physical excitation
of the global symmetry breaking vacuum, the global Higgs, can be
amongst the lightest states of a strongly-coupled UV completion of
composite Higgs scenarios.  It should couple to the Higgs and
longitudinal electroweak gauge bosons at tree-level, and to
the SM fermions proportionally to their masses.  In addition, the
global Higgs interacts at 1-loop with the SM gauge bosons via loops of
the (possibly many) fermion resonances.  This last feature is
dependent on the realization of the fermion sector.  The production of
the global Higgs and its study may thus shed some light on the UV
completion of composite Higgs models.  In the next sections we explore
in more detail the expected properties of the global Higgs in specific
scenarios.

\section{Bosonic Couplings}
\label{se:boson_couplings}

We turn now to the detailed properties of the global Higgs, starting
with its dominant tree-level interactions to bosons, which are rather
model independent.  The fermion sector will be discussed subsequently.
To be definite we will focus on the case where $G=SO(5)$ and $G' =
SO(4)=SU(2)_L\times SU(2)_R$. Here, weak isospin is identified with $SU(2)_L$, while hypercharge is embedded as 
$Y=T_R^3$ of $SU(2)_R$.\footnote{There is also a $U(1)_X$ factor, such that hypercharge is actually $Y=T_R^3+X$. Only fermions carry non-zero $X$ charge, see Tab.~\ref{tab:representations}.
}

In order to describe the vacuum fluctuations of the $SO(5)/SO(4)$
sigma-model, the $SO(5)/SO(4)$ NGBs need to be embedded into a $SO(5)$
representation.  We choose to embed them into the fundamental $\bf 5$
of $SO(5)$,\footnote{We present in App.~\ref{se:14} the embedding into
a symmetric traceless $\bf 14$ of $SO(5)$.} by defining a scalar
$\Phi$ parametrized as
\be
\Phi = U_5 \H\,, 
\ee
where $U_5$ is the $SO(5)/SO(4)$ NGB matrix and $\H$ the remaining
``radial'' degree of freedom.  The vev of $\H$ will be denoted by
$\hat f e_5$, where $e_5$ denotes a unit vector in the radial
direction.  The fluctuation $\phi$ of $\H$ around its vev, given by
\be\H=(\hat f+ \phi) e_5\,,\ee is the global Higgs.

\subsection{Self-Couplings}

Having introduced the $SO(4)$ singlet degree of freedom $\H$, a
non-trivial potential $V(\H)$ is needed to stabilize it at the
non-zero vev $\left\langle \H\right\rangle =\hat f \neq 0$ that breaks
the global symmetry down to $SO(4)$.  The potential for $\H$ is in
principle arbitrary, the only requirements being $\frac{d}{d\H}\left.
V(\H)\right|_{\H=\hat f}=0$, $\frac{d^2}{d\H^2}\left.
V(\H)\right|_{\H=\hat f}>0$.  As the global Higgs self-interactions
are irrelevant for low-energy phenomenology, it is enough to consider
the expansion of $V(\H)$ up to quartic order,
\be
V(\H) =  \frac{1}{4} \lambda \left( \H^2 - \hat{f}^2 \right)^2\,.\label{eq:GH_potential}
\ee
With this parametrization, the $\lambda$ parameter corresponds to the
quartic coupling of the global Higgs, and the global Higgs mass is
given by
\be
m_\phi = \sqrt{2\lambda} \hat{f}~.
\label{eq:GH_mass}
\ee
%

\subsection{Couplings to the Goldstone Bosons and Vector Resonances  } \label{se:GH_bosonic}

Although we have argued above that the spin-1 resonances may be
amongst the heaviest new physics states (and are therefore not the
focus of this work), their presence can still leave an imprint in the
properties of the global Higgs, which results in an additional free
parameter $f_\rho$.  We therefore present the bosonic sector, that
consists of $\H $, the pNGB's in $\Phi$ (\ie~the SM Higgs doublet) and
a complete spin-1 multiplet of $SO(5)$ (in the adjoint
representation), in addition to the elementary gauge bosons that give
rise to the SM gauge boson sector.

Quite generically, the global Higgs couples to the pNGB states in
$\Phi$ and to the various spin-1 states.  These couplings follow from
the bosonic Lagrangian (see, e.g.~\cite{Bando:1987br})
\bea
\mathcal L_{\rm bos} &=& 
 \frac{1}{2}(\nabla_\mu\H)^2 - V(\H)
+ \frac{1}{4} f^2_\rho \left( \A^A_\mu-i[ U_{5}^\dagger  D_\mu U_{ 5}]^A \right)^2 
\nonumber \\ [0.3em]
& & 
\mbox{} - \frac{1}{4g_\rho^2}(\F^A_{\mu\nu})^2
-\frac{1}{4g_0^2}(w_{\mu\nu}^a)^2 - \frac{1}{4g'^2_0}(b_{\mu\nu})^2~,
\label{Lbos}
\eea
where the covariant derivative $D_\mu$ contains the elementary gauge
fields $w_\mu$ and $b_\mu$ only.  The heavy spin-1 resonances are
denoted by $\A_\mu^A$, including both $SO(4)$ ($A=a$) and
$SO(5)/SO(4)$ ($A=\hat a$) resonances.  We have used the following
definitions
\be
U_{5}=\exp\left(\frac{\sqrt{2}\,i}{f}\,h^{\hat a}T^{\hat a}\right)~,
\qquad
\nabla_\mu \H=\left(\partial_\mu-i\,\A_\mu^{\hat a}\,T^{\hat a}\right)\H~.
\ee
In Ref.~\cite{vonGersdorff:2015fta} it was shown that this Lagrangian
can be obtained from a theory of 4-fermion interactions.  As already
pointed out there, $\mathcal L_{\rm bos}$ is in fact more general, as
it is recognized to correspond to a 2-site model Lagrangian where the
radial mode of the $SO(5)\to SO(4)$ breaking of the second site is
included.  Therefore, the couplings of $\phi$ should be fairly general
and applicable to a wide class of UV completions of models considered
in the literature.

Notice that when $f_\rho\to \infty$, which sets $\A^A_\mu = i[
U_{5}^\dagger D_\mu U_{ 5}]^A$, one can rewrite the theory as a
\textit{linear} sigma model in terms of the fiveplet $\Phi=U_5\,\H$.
For finite $f_\rho$, the couplings of $\H$ to the pNGB's deviate from
those of the linear sigma model due to the mixing with the coset
spin-1 resonances.  In order to diagonalize this mixing, we focus on
the terms quadratic in the pNGB's $h^{\hat{a}}$ and the coset
resonances $\mathcal A^{\hat a}$.  We get
\begin{multline}
\frac{1}{2}|\nabla \H|^2+\frac{1}{4}f_\rho^2
\left( \A^A_\mu-i[ U_{5}^\dagger  D_\mu U_{ 5}]^A \right)^2\\
=\frac{1}{2}(\partial_\mu\phi)^2+\frac{1}{4}(\hat f+\phi)^2(\A_\mu^{\hat a})^2+\frac{f_\rho^2}{4}\left(\A^{\hat a}_\mu+\frac{\sqrt{2}}{f}D_\mu h^{\hat a}\right)^2+\cdots
\label{eq:GBkin}
\end{multline}
To disentangle the Goldstones and the vector bosons, define the
shifted field $\B_\mu$ and choose $f$ according to
\be
\A_\mu=\B_\mu-\frac{\sqrt{2}\,f}{\hat f^2}D_\mu h\,,\qquad f^{-2}=\hat f^{-2} + f_\rho^{-2}~,
\ee
which eliminates the cross term $D_\mu h^{\hat a} \A_\mu^{\hat a} $
and renders the pNGB kinetic term canonical:
\be
\mathcal L=
\frac{1}{2} (\partial_\mu\phi)^2+\frac{1}{2}(D_\mu h^{\hat a})^2+\frac{m_a^2}{2}(\B_\mu^{\hat a})^2+\left(\frac{1}{2}\hat f\phi+\frac{1}{4}\phi^2\right)
\left(\B_\mu^{\hat a}-\frac{\sqrt{2}\,f}{\hat f^2}D_\mu h^{\hat a}\right)^2~.
\ee
The last term contains all the interactions.  In particular one has
the following interaction linear in $\phi$:
\bea
\mathcal L &\supset& \frac{f^2}{\hat f^3}\phi\, (D_\mu h^{\hat a})^2
~\equiv~ 
f_H^{-1} \mathcal O_H~,
\label{GHNGBInt}
\eea
with $\mathcal O_H = \frac{1}{2}\phi \, (D_\mu h^{\hat a})^2 = \phi \,
|D_\mu H|^2$, where $H = \frac{1}{\sqrt{2}}(h^{\hat{2}} + i
h^{\hat{1}}, h^{\hat{4}} - i h^{\hat{3}})^T$ is the SM Higgs doublet.
We therefore identify the induced coefficient for this dimension-5
interaction as $f_H^{-1}=2\, r_v\,\hat f^{-1}$, with
\bea
r_v &\equiv& \frac{f^2}{\hat{f}^2} ~=~ \frac{f_\rho^2}{f_\rho^2 + \hat{f}^2} ~=~ \frac{m_\rho^2}{m_a^2}~\leq 1.
\label{rv}
\eea
Here we have identified the mass of the spin-1 $SO(4)$ resonances
$\A_\mu^a$, given by $m_\rho^2 = \frac{1}{2} g_\rho^2 f_\rho^2$, and
of the $SO(5)/SO(4)$ resonances $\B_\mu^{\hat a}$, given by $m_a^2 =
\frac{1}{2} g_\rho^2 (f_\rho^2 + \hat{f}^2)$.

The linear sigma model result $f_H^{-1}=2\,\hat f^{-1}$ is indeed
recovered in the limit $f_\rho\to \infty$.  For finite $f_\rho$,
however, the scale suppressing the interaction in Eq.~(\ref{GHNGBInt})
can differ by order one from both the naive decay constant $\hat{f}$
which appears in other interactions (see the next section), as well as
from the Higgs decay constant, $f$, which controls the couplings of
the pNGB Higgs.  As we can see, $f_H$ sets the coupling of a single
global Higgs to both a pair of SM Higgses, as well as to pairs of W's
and Z's through their longitudinal polarizations, i.e., couplings of the form $m_V^2f_H^{-1} \phi V_\mu V^\mu$.

One notices that couplings of the global Higgs to transversely
polarized SM gauge bosons are absent at this level (they are
suppressed by a loop factor or, after including EWSB effects, by
${\cal O}(v^2/f_H^2)$).  However, they are crucial for
phenomenological studies and will be addressed later.

\section{Fermionic Couplings}
\label{se:fermion_couplings}

The couplings of the global Higgs to the fermions in the theory are
more model dependent.  First, one should notice that in the previous
section we considered the simplest possibility where there exists a
single global Higgs that, together with the four pNGB's that
constitute the SM Higgs doublet, falls into a 5 of $SO(5)$.  It may be
possible, however, that the pNGB's arise from larger $SO(5)$
representations.  Connected to this, there is significant
model-building freedom to choose the $G$-multiplets for the fermionic
resonances, the only constraint being that they contain a subset of
states with the appropriate SM quantum numbers to allow mixing and the
implementation of the partial compositeness paradigm.

We will therefore be content with describing some illustrative
possibilities and settle on a few representative benchmark scenarios,
that could be used for further phenomenological studies.  In order to
set up the framework, we will start by focusing on a simple top
sector.  We will then comment on possible variations and on the
corresponding constructions necessary for the lighter generations
(more precisely, the differences between the up-quark, down-quark and
lepton sectors).

\subsection{A Simple Top Sector}
\label{SimplestTopSector}

We are interested in the coupling of $\phi$ to fermion pairs which
arises from the $SO(5)$ symmetric Yukawa couplings, as in
Eq.~(\ref{proto}).  We will first consider a minimal top-sector,
consisting of vector-like top-partners $F$ and $S$, which transform in
the $\bf 5_\frac{2}{3}$ and $\bf 1_{\frac{2}{3}}$ of the $SO(5)\times
U(1)_X$ group respectively.  In addition, we include two elementary
fields $q^{el}_L$ and $t^{el}_R$ with the usual SM quantum numbers.
The $SO(5)$ Higgs, $\Phi$, will be assumed to transform in the
fundamental of $SO(5)$, as in the previous subsection.  We can
therefore write the Yukawa coupling \cite{vonGersdorff:2015fta}
\be
\mathcal L_{\xi_t}=-\xi_t\, \Phi^i \left(\bar F^i_LS_R^{\ } + {\rm h.c.} \right )\,,
\ee
where $\Phi=U_5\,(\hat f+\phi) \, e_5$.  One can immediately verify
that the $SO(4)$ four-plet arising from $F$ does not acquire Yukawa
couplings to $\phi$ before EWSB. We can therefore focus on the
$SU(2)_L$ singlet sector which consists of two left-handed fields,
$\left(F_L^5\,,\ S_L\right)$ as well as three right-handed fields
$\left(F_R^5\,,\ S_R,\,\ t_R^{el}\right)$.  Under the SM, these fields
are $SU(2)$ singlets with hypercharge $\frac{2}{3}$, \ie~they
transform like the right handed top quark.  There will thus be in
general one mixing angle in the left-handed sector and three in the
right-handed one.  We will simplify the discussion by decoupling one
vectorlike state [for instance $\left(S_L\,,\ t^{el}_R\right)$ or
$\left(S^{\ }_L\,,\ F^5_R\right)$], so that one is left with only one
mixing angle $s_{t_R}=\sin \alpha_{t_R}$, that rotates the remaining
two right-handed fields to the mass eigenbasis $T_R,t_R$.  The
Lagrangian of the hypercharge $\frac{2}{3}$ top states then reads
\be
\mathcal L_{1_\frac{2}{3}}=-m_T \,\bar T_L T_R -\xi_t\, \phi\left(c_{t_R}\,\bar T_L\,T_R+s_{t_R}\,\bar T_L\,t_R\right) + {\rm h.c.}
\ee
with $m_T=\xi_t\hat f/c_{t_R}$ and $t_R$ denotes the physical
right-handed top quark.  Note that neglecting electroweak breaking,
the physical top-quark does not possess Yukawa coupling to $\phi$,
only a ``mixed" one involving also the heavy top resonance.

After electroweak symmetry breaking, the physical top quark acquires
also a Yukawa coupling to $\phi$, which is universally given by
\be
\mathcal L_{\phi \bar tt}=-\frac{m_t}{\hat f}\phi\, \bar t\, t~.
\ee 
In the following we will neglect EWSB effects, since they are a small
perturbation for the physics at $\hat{f}$.

\subsection{Other Embeddings and Light Quarks}
\label{lightquarks}

The above choice of top partners is by no means unique.  There exist
many choices for the representations of the top (and other fermion)
partners.  As already stated, the paradigm of partial compositeness
simply requires that all SM fermions appear in these representations
at least once (such that mixing with the elementary states can take
place), and that at least one of each kind appear in the
$SO(5)$-invariant Yukawa couplings.  The typical representations
considered in the literature (see, e.g.~\cite{Carena:2014ria}) are the
{\bf 1, 5, 10} or {\bf 14} of $SO(5)$.  They will be denoted by $S$,
$F$, $A$, $B$ respectively.  Their decompositions are detailed in
Table~\ref{tab:representations}.

\begin{table}[t]
\begin{center}
\begin{tabular}{|c|c|c|}
\hline 
\rule{0mm}{5mm}
$SO(5) \times U(1)_X$				& $SO(4) \times U(1)_X$			& $SU(2)_L \times U(1)_Y$
\\ [0.4em]
\hline 
\hline 
$1_{\frac{2}{3}}$	& $1_\frac{2}{3}$ 	& $1_\frac{2}{3}$
\\ [0.4em]
\hline
$5_\frac{2}{3}$		& $1_\frac{2}{3}+4_\frac{2}{3}$
										& $1_\frac{2}{3}+(2_\frac{1}{6}+2_\frac{7}{6})$
\\ [0.4em]
\hline
$5_{-\frac{1}{3}}$	& $ 1_{-\frac{1}{3}}+4_{-\frac{1}{3}} $
										& $1_{-\frac{1}{3}}+(2_{-\frac{5}{6}}+2_{\frac{1}{6}})$
\\ [0.4em]
\hline
$5_{-1}$	& $ 1_{-1}+4_{-1} $
										& $1_{-1}+(2_{-\frac{3}{2}}+2_{\frac{1}{2}})$
\\ [0.4em]
\hline
$10_{\frac{2}{3}}$	& $ 4_\frac{2}{3}+6_\frac{2}{3}$
								& $(2_\frac{1}{6}+2_\frac{7}{6})+(1_{-\frac{1}{3}}+1_{\frac{2}{3}}+1_\frac{5}{3}+3_\frac{2}{3}) $
\\ [0.4em]
\hline
$14_\frac{2}{3}$	& $1_\frac{2}{3}+4_{\frac{2}{3}}+9_{\frac{2}{3}}$
								& $1_{\frac{2}{3}}+(2_\frac{1}{6}+2_\frac{7}{6})+(3_{-\frac{1}{3}}+3_{\frac{2}{3}}+3_\frac{5}{3})$
\\ [0.4em]
\hline 
\end{tabular}
\caption{Decomposition of the smallest $SO(5)\times U(1)_X$
representations under both the custodial $SO(4)$ and the SM $SU(2)_L
\times U(1)_Y$.
\label{tab:representations}}
\end{center}
\end{table}

Instead of working out in detail other possible top sectors, we will
move on to describe various possibilities that can also be applied to
the composite states that partner with the light SM fermions.  We
start by noticing that not all combinations of fermion partners allow
for simple renormalizable Yukawa couplings with the $SO(5)$ breaking
field in the fundamental.  For instance, choosing two quark partners
$F$ and $F'$, one either needs to resort to a $SO(5)$ breaking field
in the $\bf 14$, or to nonrenormalizable Yukawa couplings:
\be
\mathcal L^{\rm dim\, 4}_\xi=-\xi\, \bar F^i \Psi_{ij} F'^j~,
\qquad {\rm or} \qquad  
\mathcal L^{\rm dim\, 5}_\xi=-\frac{\xi}{\hat f}\, \bar F^i \Phi^i F'^j \Phi^j~.
\label{YukSO5}
\ee
In the case that the pNGB's arise from a $\bf 14$, we define the
global Higgs as the mode in the $SO(4)$ singlet direction
$\frac{1}{\sqrt{20}}{\rm diag}(1,1,1,1,-4)$.  The two choices above
lead to different couplings between $\phi$ and the various $SO(4)$
representations.  The possible $SO(4)$ representations from the
decompositions in Table~\ref{tab:representations} are the $\bf 1,\ \bf
4,\ 6,\ $ and $\bf 9$ and are assumed to be canonically normalized.
We denote them by $s$, $f$, $a$ and $b$, respectively.  The $SO(5)$
symmetric proto-Yukawa couplings induce Yukawa interactions with the
global Higgs, e.g.~
\be
\bar F\Psi F' =\frac{2}{\sqrt{5}}  \phi\, \bar s s' +\frac{1}{2\sqrt 5}\phi\, \bar f\!f'~.
\ee
We denote these weight factors by $w_i$, such that the $SO(4)$
representation labeled by $i$ couples to the global Higgs with Yukawa
coupling
\be
\xi_{U,i}=w_i\xi_U~,\qquad \xi_{D,i}=w_i\xi_D~,\qquad \xi_{E,i}=w_i\xi_E~.
\label{weights}
\ee
The various possible $SO(5)$ and $SO(4)$ symmetric Yukawa couplings
for the above representations and the respective factors $w_i$ are
summarized in Tables~\ref{tab:yukawas4} and \ref{tab:yukawas5} (see
App.~\ref{app:Yukawas} for further details).  We see, in particular,
that the number of fermion states that couple to the global Higgs
depends very much on the assumed representation of both the scalar and
the fermions.

We will assume all masses and couplings in the fermionic Lagrangian to
be real for definiteness.\footnote{In realistic scenarios the phases
are constrained by CP violation.  Constructing a fully realistic
flavor sector is not the goal of this work.} Notice that we can write
two independent Yukawas of the type $\Phi \bar Q U$ and $\Phi \bar Q
\gamma^5 U$.  We find it more convenient to switch to the two
operators $\Phi \bar QP_{R}U$ and $\Phi\bar Q P_L U$, whose
coefficients we will generally denote by $\xi$ and $\xi'$
respectively.

\begin{table}[t]
\begin{center}
\begin{tabular}{|c||ccc|ccccc|}
\hline
\rule{0mm}{5mm}
proto-Yukawa 	&$\bar F\Phi S'$	&$\bar FA' \Phi$	&$\bar F B' \Phi$	&
	$\bar F\Psi F'$	& $\bar S\tr\Psi B'$	& $\tr \bar A\Psi A'$ &$\tr \bar B\Psi B'$	&$\tr \bar B\Psi A'$
\\ [0.4em]
\hline
\hline
$\phi\, \bar s s'$  &1&$-$&$\frac{2}{\sqrt{5}}$&
	$\frac{2}{\sqrt 5}$			& 1					& $-$				& $\frac{3}{2\sqrt 5}$			& $-$
\\ [0.4em]	
$\phi\,\bar f\! f'$  &$-$&$\frac{1}{\sqrt 2}$&$\frac{1}{\sqrt 2}$&
	$\frac{1}{2\sqrt 5}$			& $-$					& $\frac{3}{4\sqrt 5}$				& $\frac{3}{4\sqrt 5}$				& $\frac{\sqrt 5}{ 4}$
\\ [0.4em]	
$\phi\, \bar a a'$  &$-$&$-$&$-$&
	$-$			& $-$					& $\frac{1}{2\sqrt 5}$				& $-$				& $-$
\\ [0.4em]	
$\phi\, \bar b b'$  &$-$&$-$&$-$&
	$-$			& $-$				& $-$				& $\frac{1}{2\sqrt 5}$				& $-$
\\ [0.4em]	
\hline
\end{tabular}
\caption{The $SO(4)$ multiplets that couple to the global Higgs, for
various choices of the dimension-4 proto-Yukawa interactions (as
specified in the first row).  The entries give the weight factors
$w_i$ as defined in Eq.(\ref{weights}).  We use the notation $S
\leftrightarrow {\bf 1}$, $F \leftrightarrow {\bf 5}$, $A
\leftrightarrow {\bf 10}$ and $B \leftrightarrow {\bf 14}$ to indicate
the various fermionic $SO(5)$ representations considered, as well as
$s \leftrightarrow {\bf 1}$, $f \leftrightarrow {\bf 4}$, $a
\leftrightarrow {\bf 6}$ and $b \leftrightarrow {\bf 9}$ for the
$SO(4)$ representations.  $\Phi$ is understood as a fiveplet of
$SO(5)$, while the matrix $\Psi$ is to be interpreted as the {\bf 14}
representation of $SO(5)$.
\label{tab:yukawas4}}
\end{center}
\end{table}
\begin{table}[t]
\begin{center}
\begin{tabular}{|c||ccccc|}
\hline
\rule{0mm}{5mm}
proto-Yukawa 	&
	$\bar F\Phi\, \Phi^\dagger\!F'$	& $\bar S\, \Phi^\dagger  B'\Phi$	& $ \Phi^\dagger\bar A  A'\Phi$ &$\Phi^\dagger\bar BB' \Phi$	&$ \Phi^\dagger\bar B A'\Phi$
\\ [0.4em]
\hline
\hline
$\phi\ \bar s s'$	&2		& $\frac{4}{\sqrt 5}$					& $-$					&$\frac{8}{5}$ 		&$-$
\\ [0.4em]
$\phi\,\bar f\! f'$	&$-$		& $-$					& 1					&1 		&1
\\ [0.4em]
\hline
\end{tabular}
\caption{Same as Table \ref{tab:yukawas4}, but for the dimension-5
proto-Yukawa interactions.
\label{tab:yukawas5}}
\end{center}
\end{table}

For the lighter up-type quarks one can mimic the construction
described in more detail for the top quark sector in
Subsection~\ref{SimplestTopSector}, which falls in the ``$\bar{F} \Phi
S$" category of Table~\ref{tab:yukawas4}.  Alternatively, one can use
the less minimal variant that replaces the ${\bf 1}_{\frac{2}{3}}$
with the ${\bf 14}_{\frac{2}{3}}$ of $SO(5) \times U(1)_X$,
corresponding to the ``$\bar{F} B \Phi$" category.  For the bottom
sector (specifically the $b_R$) one can see from
Table~\ref{tab:representations} that the candidate representations are
the ${\bf 5}_{-\frac{1}{3}}$ and the ${\bf 10}_{\frac{2}{3}}$.  The
first case requires the introduction of an additional composite ${\bf
5}_{-\frac{1}{3}}$ partner of the $(t_L,b_L)$ doublet, in order to be
able to write down the $\bar{F} \Psi F$ category of
Table~\ref{tab:yukawas4} for the bottom sector.  This would be in
addition to the composite ${\bf 5}_{\frac{2}{3}}$ associated with the
top sector, which is also a composite partner of the $(t_L,b_L)$
doublet.  Thus, the choice of a composite ${\bf 5}_{-\frac{1}{3}}$
that partners with $b_R$ leads to a rather non-minimal scenario.  If
one insists on dim-4 proto-Yukawa interactions, such that $\Psi$,
which hosts the global Higgs, would transform in the ${\bf 14}$ of
$SO(5)$, one increases even more the level of complexity.  The second
option is more minimal in comparison: with the $b_R$ composite partner
transforming in the ${\bf 10}_{\frac{2}{3}}$ of $SO(5) \times U(1)_X$
one can write a dim-4 proto-Yukawa interaction using the global Higgs
arising from a ${\bf 5}$ (denoted by $\Phi$ before), and without
enlarging the top sector.  We will therefore take this second case,
replicated for all the down-type quarks as a reference example.

\subsection{Benchmark Models}
\label{benchmarks}

We now define a set of benchmark scenarios in order to illustrate the
typical embeddings of the global Higgs in composite Higgs models.  In
a forthcoming publication, a collider analysis will be
carried out for these scenarios \cite{wip}.  \\[-0.7em]

\textit{\underline{Quark Benchmarks}}:
%
%
%
%
%
\begin{table}[h]
\begin{center}
\begin{tabular}{lllc}
$\bullet$~~MCHM$_{5,1,10}$: 
& $(Q_i, U_i, D_i) = ({\bf 5}_{\frac{2}{3}}, {\bf 1}_{\frac{2}{3}}, {\bf 10}_{\frac{2}{3}})$~,
& $\phi \subset {\bf 5_0}$~,
& \hspace{2.9cm}
\\ [0.5em]
$\bullet$~~MCHM$_{5,14,10}$: 
& $(Q_i, U_i, D_i) = ({\bf 5}_{\frac{2}{3}}, {\bf 14}_{\frac{2}{3}}, {\bf 10}_{\frac{2}{3}})$~,
& $\phi \subset {\bf 5_0}$~,
\\ [0.5em]
$\bullet$~~MCHM$_{14,14,10}$: 
& $(Q_i, U_i, D_i) = ({\bf 14}_{\frac{2}{3}}, {\bf 14}_{\frac{2}{3}}, {\bf 10}_{\frac{2}{3}})$~, 
& $\phi \subset {\bf 14_0}$~.
\end{tabular}
\end{center}
\end{table}

\vspace{-5mm}
\noindent
The first two models require the global Higgs to be embedded in the
{\bf 5} representation, while for the third one we chose the {\bf 14}.
The last model uses the ``${\rm tr} \, \bar{B} \Psi B$" proto-Yukawa
structure for the up sector, and the ``${\rm tr} \, \bar{B} \Psi A$"
proto-Yukawa structure for the down sector, following the notation of
Table~\ref{tab:yukawas4}.  These three benchmark models are understood
to be characterized by order one proto-Yukawa couplings (we will be
more precise in Sec.~\ref{se:perturbativity}) and hierarchical mixing
angles (see the discussion after Eqs.~(\ref{proto}) and (\ref{mix}) in
Section~\ref{se:scenarios}).

As already mentioned in Section~\ref{se:scenarios} we will also
consider the scenario with hierarchical Yukawas and order one mixing
angles, as a ``most minimal" example where the global Higgs properties
are only sensitive to the top sector:
\begin{itemize}

\item MCHM$_{5,1}$: ~~$(Q_3, U_3) = ({\bf 5}_{\frac{2}{3}}, {\bf
1}_{\frac{2}{3}})$.  The representations of the composite partners for
the SM fermions other than $q_L = (t_L, b_L)$ and $t_R$ need not be
specified in this case since they play a negligible role.  The global
Higgs is in a ${\bf 5_0}$.

\end{itemize}

\textit{\underline{Lepton Benchmarks}}:
\vspace{2mm}

The lepton sector can potentially play a role in the coupling of the
global Higgs to two photons, and as for the case of quarks vis-\`a-vis
the two-gluons amplitude, they can introduce additional
model-dependence.  We therefore fix two benchmark scenarios in the
leptonic sector, that apply for each of the four quark sector
scenarios defined above:
\begin{itemize}

\item A leptonic anarchic scenario with~$(L_i, E_i) = ({\bf 5}_{-1},
{\bf 1}_{-1})$, which falls in the ``$\bar{F} \Phi S$" category of
Table~\ref{tab:yukawas4}.

\item A non-anarchic scenario analogous to the MCHM$_{5,1}$ above,
where all the composite lepton proto-Yukawa couplings are small, and
therefore the heavy leptonic states have a minimal impact on the
phenomenology of the global Higgs.  In this case, the representations
of the composite partners of the SM leptons need not be specified.

\end{itemize}

As we will see, this will illustrate that the impact of the leptonic
sector can be relatively minor.  Notice also that here we will remain
agnostic about the composite states related to the neutrino sector,
and assume, conservatively, that they do not contribute.

\section{Effective One-loop Couplings to the SM Gauge Bosons}
\label{se:oneloop}

As mentioned before, the global Higgs couplings to massless gauge
bosons such as the gluon and the photon as well as the couplings to
the transverse polarizations of the electroweak gauge bosons are
induced by one-loop processes which are sensitive to the details of
the particles running in the loop.  These couplings are very important
for phenomenological studies and in this section we estimate them in
the different benchmark models defined in the previous section.

The effective one-loop coupling of the global Higgs to a gluon pair
proceeds in complete analogy to the SM calculation.  The result can be
encoded into the dimension-5 effective term
\bea
{\cal L}^{\rm eff}_{\phi gg} &=& - \frac{\alpha_s N_{\phi gg}}{12 \pi \hat{f}} \, \phi \, G^a_{\mu\nu} G^{\mu\nu}_a~,
\label{Lphigg}
\eea
where
\bea
N_{\phi gg} = \frac{3}{4}\hat f \sum_i \frac{M_i'}{M_i} \, A_{1/2}\!\left( \frac{m^2_\phi}{4M^2_{i}} \right)~,
\label{Nphigg}
\eea
and the sum runs over all the quark states of mass $M_i$ that couple
to the global Higgs $\phi$ with Yukawa strength $M_i'=\partial_\phi
M_i(\phi)|_{\phi=0}$.\footnote{For simplicity, and because it is a
good approximation, we will neglect EWSB effects.} The $A_{1/2}(\tau)$
is the standard loop function, which is given in
Appendix~\ref{loopfunctions}.  It saturates to 4/3 in the limit that
the fermion is heavy compared to $m_\phi$, and vanishes in the
opposite limit.

The coupling of the global Higgs to the EW gauge bosons is similarly
given by
\bea
- \frac{\alpha}{s^2_W} \, \frac{N_{\phi WW}}{8 \pi \hat{f}} \, \phi \, W^i_{\mu\nu} W^{\mu\nu\, i}
- \frac{\alpha}{c^2_W} \, \frac{N_{\phi BB}}{8 \pi \hat{f}} \, \phi \, B_{\mu\nu} B^{\mu\nu}~,
\label{LphiWWBB}
\eea
where we can write $N_{\phi WW} = N_{\phi \gamma\gamma} - N_{\phi BB}$
with
\bea
N_{\phi \gamma\gamma} &=& \frac{m^2_a - m_\rho^2}{m^2_{a}} \, A_{1}\!\left( \frac{m^2_\phi}{4m^2_{a}} \right) + \hat f \sum_i 
\frac{M_i'}{M_i}\, N_c Q_i^2 A_{1/2}\!\left( \frac{m^2_\phi}{4M^2_{i}} \right)~,
\label{Nphigaga}
\\ [0.4em]
N_{\phi BB} &=& \hat f\sum_i 
\frac{M_i'}{M_i}N_c Y_i^2 A_{1/2}\!\left( \frac{m^2_\phi}{4M^2_{i}} \right)~,
\label{NphiBB}
\eea
and $N_c = 3\,(1)$ is the number of colors for quarks (leptons).  One
then gets the couplings to $\gamma\gamma$, $ZZ$, $\gamma Z$ and
$W^+W^-$:
\bea
- \frac{\alpha}{8 \pi \hat{f}}
\left(
N_{\phi \gamma\gamma} \, \phi \, F_{\mu\nu} F^{\mu\nu}
+ \frac{N_{\phi ZZ}}{s^2_W c^2_W} \, \phi \, Z_{\mu\nu} Z^{\mu\nu}
+ \frac{2N_{\phi Z\gamma}}{s_W c_W} \, \phi \, F_{\mu\nu} Z^{\mu\nu}
+ \frac{2N_{\phi WW}}{s^2_W} \, \phi \, W^+_{\mu\nu} W^{-\mu\nu}
\right)~,
\nonumber \\
\label{Lphigaga}
\eea
where
\bea
N_{\phi ZZ} = c^4_W N_{\phi WW} + s^4_W N_{\phi BB}~,
\hspace{7mm}
N_{\phi \gamma Z} = N_{\phi WW} - s^2_W N_{\phi \gamma\gamma}~.
\eea
The first term in Eq.~(\ref{Nphigaga}) corresponds to the charged pair
of $SO(5)/SO(4)$ gauge bosons that receive a contribution $\sqrt{m^2_a
- m_\rho^2}$ to their mass from the breaking at $\hat{f}$.  The
well-known loop function $A_{1}(\tau)$ (see
Appendix~\ref{loopfunctions}) reaches the asymptotic value $-7$ when
the spin-1 resonance is much heavier than the global Higgs.  Notice
that the spin-1 contribution is completely parallel to the one from
the charged $W$'s in the SM Higgs case, with the pair of heavy charged
vector fields playing the role of the SM $W^{\pm}$.  Here, however, we
expect to be much closer to the saturation limit of the loop function,
since the spin-1 resonances are taken to be heavy (see the discussion
in Section~\ref{se:scenarios}).  The second term in
Eq.~(\ref{Nphigaga}), as well as Eq.~(\ref{NphiBB}), includes the
contribution from all the fermions that couple to the global Higgs,
including quark and lepton fields.

In order to estimate the multiplicity factors $N_{\phi XX}$ with $X =
g, \gamma, B$ for each of the benchmark scenarios defined in the
previous subsection, we assume that the fermions have a common
vector-like mass, $M_Q=M_U=M_D=M_\psi$.  If the vectorlike mass
dominates over the global symmetry breaking effects, and assuming also
small mixing with the elementary sector, the fermions are
approximately degenerate in mass with $M_i\sim M_\psi $.\footnote{As
we will detail in the accompanying work~\cite{wip}, our analytic
expressions for the loop processes can be quite effectively used even
when these assumptions are not fulfilled, by using
Eq.~(\ref{eq:coupgg}) to \textit{define} an effective mass scale
$M_\psi$ (provided it does not vanish; see next footnote).  The $\phi\gamma\gamma$ and $\phi BB$ processes can
similarly be used to define effective scales via
Eq.~(\ref{eq:coupgamgam}) and the analogous equation with hypercharge
weighting.  However, in the bulk of the parameter space of each model,
the three scales are quite similar and, therefore, to a good
approximation, one can reduce the model dependence to a single
parameter, that one can characterize as ``the scale of spin-1/2
resonances".} We can then factor out a common loop function
$A_{\frac{1}{2}}(m_\phi^2/4M^2_\psi)$ and compute the sum as~\footnote{Note that for this result to be non-vanishing both $\xi$ and $\xi'$ must be non-zero. We recall that in the presence of the ``wrong-chirality" structure with coefficient $\xi'$, the SM Higgs potential may acquire a log sensitivity to the compositeness scale $\Lambda$. This mild dependence is not necessarily a problem. If one has a situation with a vanishing $\xi'$, the sum in Eq.~(\ref{eq:coupgg}) is proportional to the elementary-composite mixing, which we have ignored in the derivation.}
\be
 \sum_i \frac{M_i'}{M_i}= \frac{d(\det \mathcal M)/d\hat{f}}{\det \mathcal M}
\approx -2\frac{\hat f}{M_\psi^2}\left(\bar N^U_{\phi g g}\tr \xi_U'\xi_U^T +\bar N^D_{\phi gg }\tr \xi'_D\xi_D^T \right)~,
\label{eq:coupgg}
\ee
where the remaining traces are over the 3 generations, and the $\bar
N^{U,D}_{\phi gg}$ are the sums over the $SO(4)$ multiplicities $N_i$,
weighted by the factors $w_i^2$, where the $w_i$ are given in Table~\ref{tab:yukawas4}.  The $\bar N^{U,D}_{\phi gg}$ are summarized in
Table \ref{tab:multi}.  Analogously, we can obtain
\be
\sum_i \frac{M_i'}{M_i} N_c Q_i^2
\approx -2\frac{\hat f}{M_\psi^2}\left(3 \bar N^U_{\phi\gamma\gamma}\tr \xi'_U\xi_U^T + 3 \bar N^D_{\phi\gamma\gamma}\tr \xi'_D\xi_D^T 
+\bar N^E_{\phi\gamma\gamma} \tr \xi'_E\xi_E^T
\right)~,
\label{eq:coupgamgam}
\ee
where $ \bar N^{U,D,E}_{\phi \gamma\gamma}$ are the sums over the
$SO(4)$ multiplicities, weighted by the charges $Q_i^2$ and the
factors $w_i^2$ as before.  The factors $\bar N^{U,D,E}_{\phi BB}$ for
the hypercharge are defined analogously.  The charges and hypercharges
can be read off from Table~\ref{tab:representations} for each
benchmark model.  It should be noted that the tensor couplings of the electroweak gauge bosons are expected to compete with the longitudinal couplings (arising from the operator $\mathcal O_H$ at tree level) only for large Yukawa couplings and multiplicities. Moreover, we point out that the tree and loop level couplings have a different scaling with $\hat f$ or equivalently, for fixed Global Higgs mass,  with the quartic coupling $\lambda$.
The typical size of the Yukawa and quartic interactions will be
estimated in the next section.

One may wonder if higher order (finite effects) could give large
corrections to the 1-loop results, given the large multiplicities
involved.  One can see, however, that the higher order corrections
involving additional heavy fermion loops enter only at 3-loop order
and would not be expected to give a large effect.  Rather, we expect
the higher-order corrections to be dominated by QCD, very much as in
the SM. While a more precise treatment would include the QCD
$K$-factors, to be on the conservative side, we will not include any
such corrections.  One should, however, keep in mind that they will
give an additional enhancement to the rates involving two gluons or
two photons.

\begin{table}[t]
\begin{center}
\begin{tabular}{|c||cc|ccc|ccc|}
\hline 
\rule{0mm}{5mm}
Benchmark	& $\bar{N}^U_{\phi gg}$ 			& $\bar{N}^D_{\phi gg}$ 	
			& $\bar{N}^U_{\phi \gamma\gamma}$	& $\bar{N}^D_{\phi \gamma\gamma}$ & $\bar{N}^E_{\phi \gamma\gamma}$ 	
			& $\bar{N}^U_{\phi BB}$ 			& $\bar{N}^D_{\phi BB}$ & $\bar{N}^E_{\phi BB}$ 
\\ [0.1em]
\hline 
\hline 
\rule{0mm}{5mm}
MCHM$_{5,1,10}$	& 1	&2	&$\frac{4}{9}$	& $\frac{17}{9}$ &		1	& $\frac{4}{9}$ & $\frac{25}{18}$ &1
\\ [0.4em]
\hline
\rule{0mm}{5mm}
MCHM$_{5,14,10}$& $\frac{14}{5}$	& 2	& $\frac{101}{45}$	& $\frac{17}{9}$			&1	& $\frac{157}{90}$	& $\frac{25}{18}$ &1
\\ [0.4em]
\hline
\rule{0mm}{5mm}
MCHM$_{14,14,10}$&$\frac{27}{20}$	& $\frac{5}{4}$	&	 $\frac{57}{40}$	&$\frac{85}{72}$ &	1		& $\frac{81}{80}$ &$\frac{125}{144}$ &1
\\ [0.4em]
\hline 
\rule{0mm}{5mm}
MCHM$_{5,1}$	& 1	&$-$	&$\frac{4}{9}$	& $-$ & $-$	& $\frac{4}{9}$ & $-$ & $-$
\\ [0.4em]
\hline 
\end{tabular}
\caption{ Fermionic multiplicity factors entering the effective
couplings of the global Higgs to two gluons or two EW gauge bosons,
given in Eqs~\eqref{eq:coupgg} and \eqref{eq:coupgamgam}.
\label{tab:multi}}
\end{center}
\end{table}
%

\section{Running Couplings in the Composite Sector}
\label{se:perturbativity}

In the presence of large $SO(5)$ matter representations such as $\bf
10$ and $\bf 14$ (in particular when repeated for all 3 generations)
the RG running of the Yukawa couplings and of the global Higgs quartic
coupling $\lambda$ from the compositeness scale $\Lambda$ down to the
global Higgs mass scale $m_\phi$ has to be taken into account.  We
will see below that the beta function of the Yukawa couplings is
always positive because of loops of the global Higgs.  This implies
that the Yukawa couplings develop a Landau pole at relatively low
energies, and thus that the strong dynamics develops at a scale
$\Lambda$ not far above the global Higgs mass.  We shall identify this
strong coupling scale with the compositeness scale.  Below the strong
coupling scale, couplings are expected to quickly decrease, so that
the composite states can be described as well-defined propagating
states.\footnote{ We must notice that the masses of vector resonances
may in principle be higher than $\Lambda$, which means they cannot be
described consistently by the theory.  However, the imprint of these
resonances on the global Higgs properties is only characterized by
$r_v=f/\hat f$, and thus does not depend on the resonance masses.  The
kinetic terms of the vector resonances can be consistently sent to
zero by taking $g_\rho\rightarrow\infty$.  In this limit the linear
sigma model with gauge fields is strictly equivalent to a non-linear
sigma model \cite{Bando:1987br}, in which no physical particle is
present above $\Lambda$.}

We work at leading order in large multiplicities, and at 1-loop order.
It turns out that the running of the Yukawa couplings is dominated by
the wave-function renormalization of the global Higgs, and hence can
be expressed in terms of \footnote{We include here only the quark
states (which overwhelm the contribution from the lepton states).  }
\be
\xi_{\rm eff}^2=4 N_c \biggl( N^U \left[ \tr\xi_U\xi_U^T + \tr\xi'_U\xi'^T_U\right]+
N^D \left[ \tr\xi_D\xi_D^T + \tr\xi'_D\xi'^T_D\right]   \biggr )~,
\ee
where $N^U$ and $N^D$ are the multiplicities of the $SO(4)$
representations, weighted by the group-theoretical factors, and hence
they coincide with the quantities encountered in the loop expressions
for the $\phi gg$ coupling
\be
N^U=\bar N^U_{\phi gg}\,,\qquad N^D=\bar N^D_{\phi gg}\,,
\ee 
which for our various scenarios were given in Table \ref{tab:multi}.
The RG equation for $\xi_{\rm eff}$ reads
\bea
\mu\frac{d\xi^2_{\rm eff}}{d\mu} &\approx& \frac{\xi_{\rm eff}^4}{16\pi^2}~,
\label{eq:xieff}
\eea
The term above arises from the global Higgs wavefunction
renormalization, when neglecting subdominant (\ie~not enhanced by
multiplicities) terms coming from vertex and fermion wavefunction
renormalization.\footnote{We have checked that accounting for such
subleading effects the resulting corrections are indeed negligible for
our purposes of estimating the Yukawa couplings at the scale $m_\phi$.
The only exception is the minimal model MCHM$_{5,1}$, for which we
have included the subleading terms to obtain our numbers (but still
neglecting the lepton sector).} Similarly, we have neglected the
effects of gauge couplings which would also induce a differential
running between the different $\xi_i$.  It is also worth noting that,
at 1-loop order, $\lambda$ does not enter into the RG equation for
$\xi_{\rm eff}$.

\begin{table}[t]
\begin{center}
\begin{tabular}{|c||c|c||c|c|c||c|}
\hline 
\rule{0mm}{5mm}
Benchmark	& $\xi^2/\xi_{\rm eff}^2$ &$\xi (\mu=m_\phi)$&  $\epsilon $& $\lambda_{\rm min}$&$\lambda_{\rm max}$ & $\lambda/\xi^2|_{\rm fix}$
\\ [0.1em]
\hline 
\hline 
\rule{0mm}{5mm}
MCHM$_{5,1,10}$ &$\frac{1}{216}$& 0.6 (0.5) &   $\frac{1}{162}$    & 0.24 & 2.6 &1.17
\\ [0.4em]
\hline
\rule{0mm}{5mm}
MCHM$_{5,14,10}$ &  $\frac{5}{1728}$ & 0.5 (0.4) &    $\frac{11}{3456}$      & 0.12 & 2.5 &1.02
\\ [0.4em]
\hline
\rule{0mm}{5mm}
MCHM$_{14,14,10}$  & $\frac{5}{936}$ & 0.6 (0.5) & $-$ & $-$ &$-$ & $-$
\\ [0.4em]
\hline
\rule{0mm}{5mm}
MCHM$_{5,1}$		& $\frac{1}{24}$ & 1.6 (1.2) &    $\frac{1}{12}$     & 2.3 & 3.9 & 1.11
\\ [0.4em]
\hline 
\end{tabular}
\caption{ Yukawa couplings and scalar self-couplings in our various
benchmark scenarios.  See text for details.
\label{tab:couplings}}
\end{center}
\end{table}

We now assume that $\xi_{\rm eff}$ reaches a value of order $4\pi$ at
the compositeness scale $\Lambda \approx 3 m_\phi$.  This hierarchy is
somewhat arbitrarily chosen to indicate a gap between $m_\phi$ and
$\Lambda$ without taking it so large that an extreme tuning would be
involved.  With this boundary condition, we find that at the scale
$m_\phi$, the coupling $\xi_{\rm eff}$ is
\be
\xi_{\rm eff}(m_\phi)\approx 8.7 \,.
\ee
To continue further, we make the additional assumption that all the
relevant Yukawa couplings are similar.  Setting them equal ($\xi =
\xi_{U,D}=\xi_{U,D}'$) we find at the scale $m_\phi$ the values
reported in the third column of Table \ref{tab:couplings}.  The number
in parenthesis corresponds to taking $\Lambda \approx 10 m_\phi$, and
is included only for comparison.  Under our simplifying assumptions,
these are the relevant couplings when computing finite effects, such
as the loop induced couplings which are dominated by momenta of order
$m_\phi$.\footnote{We note that reproducing the top quark mass may
require taking a slightly larger $\xi_t$.  Since other couplings may
be slightly smaller, we regard our numerical estimates as representing
an average that characterizes the overall combined effect of the
multiplicity of states.}

We now turn to the quartic interaction of the global Higgs.  In order
to estimate values for $\lambda$ at $\mu = m_\phi$, we consider its
1-loop RGE, assuming our estimates for $\xi$ in Table
\ref{tab:couplings}.  The 1-loop RG equation of the quartic is given
by \footnote{For MCHM$_{14,14,10}$, we would have to consider the
simultaneous running of two quartic couplings (see App.~\ref{se:14}).}
\be
\mu\frac{d\lambda}{d\mu} \approx \frac{1}{16\pi^2}(26\lambda^2 +2\lambda \xi^2_{\rm eff} - \epsilon\, \xi^4_{\rm eff})~,
\ee
where the value of $\epsilon$ is suppressed by the multiplicities and
is reported in Table \ref{tab:couplings}.  We can see there are
actually two distinct possibilities: if $\lambda$ is sufficiently
small, it is driven to a negative value at $\Lambda$ as a result of
the renormalization by the Yukawa interactions.  Above a certain
threshold, it is driven instead to a Landau pole at $\Lambda$.  This
is similar to the well-known situation for the Higgs quartic coupling
in the SM. These two limit cases can be taken to define an upper and a
lower bound for the value of $\lambda$ at $\mu=m_\phi$.  Below we
denote these extreme values by $\lambda_{\rm min}$ and $\lambda_{\rm
max}$ .

We find that for $\Lambda = 3 m_\phi$, the quartic coupling is driven
negative when $\lambda \approx 0.24 $ at $\mu = m_\phi$ in the
MCHM$_{5,1,10}$.  For the MCHM$_{5,1}$ this value is $2.3$.  It is
larger because multiplicities are smaller in this scenario.  On the
opposite end, the maximum value of the quartic at $\mu = \Lambda$ is
given by naive dimensional analysis and is $\lambda = (4\pi)^2/3!$,
where $\xi_{\rm eff} \equiv \sqrt{N} \xi \sim 4\pi$ and $N \equiv
\sum_i 4 N_c N_i$.\footnote{The 3!  is the combinatorial factor that
we did not factor out in our definition of the quartic coupling in
Eq.~(\ref{eq:GH_potential}).} From these values at the strong coupling
scale one obtains $\lambda(\mu = m_\phi) = \lambda_{\rm max} \approx
2.6$ for the MCHM$_{5,1,10}$.  For the MCHM$_{5,1}$, taking into
account subleading corrections, one has that $\lambda(\mu = m_\phi) =
\lambda_{\rm max} \approx 3.9$.  Thus, the values of $\lambda$ are in a
even narrower range in that case.  The ranges for $\lambda$ are
summarized in Table \ref{tab:couplings}.

In connection to this, we point out that the ratio $\lambda/\xi^2$
displays a (quasi) IR fixed point which is also shown in
Table~\ref{tab:couplings}, as discussed in
\cite{vonGersdorff:2015fta}.\footnote{These fixed point values can
decrease somewhat after including the effects of the gauge
interactions.} While for the MCHM$_{5,1}$ the IR fixed point is
approached sufficiently fast, for the other models the running over
three e-folds is not sufficient to come close to the fixed point.  As
a result, we must accept an intrinsic degree of uncertainty in the
coupling $\lambda$ at $\mu = m_\phi$ in the large multiplicity models,
due to the underlying strong dynamics.  Based on the above
considerations we will allow $\lambda$ to take values in the range
$[\xi^2, \lambda_{\rm max}]$, where $\xi^2$ (which is somewhat above 
$\lambda_{\rm min}$) and $\lambda_{\rm max}$ can be
read from Table \ref{tab:couplings}.\footnote{Comparing the
MCHM$_{5,1,10}$ and MCHM$_{5,14,10}$ the slightly different lower
limits in the range for $\lambda$ are roughly consistent with the
slightly different fixed point values.  For the MCHM$_{5,1}$, on the
other hand, the uncertainty in $\lambda$ is narrower than the assumed
range, since the IR fixed point is approached more quickly.} In most
of this range, $\lambda$ is driven to its strong coupling (NDA) value
near $\Lambda = 3 m_\phi$.  Only in the vicinity of the lower limit
can $\lambda$ stay perturbative when $\mu \sim \Lambda$.  Notice also
that the quartic coupling at $\mu = m_\phi$ is always well below its
strong coupling value, as estimated by NDA.

These estimates of the couplings in the composite sector at the global
Higgs mass scale allows for more precise predictions of the global
Higgs properties.  In particular, they allow one to estimate the
one-loop effective couplings in a given scenario, and to tie the
global Higgs mass to the $SO(5)$ breaking scale $\hat f$.

\section{Decays}\label{se:decays}

The decay width of the global Higgs into SM fermions is universally given by 
\be
\Gamma_{\phi\rightarrow f \bar f } = N_c\frac{m_f^2}{8\pi \hat f ^2}\, m_\phi~,
\,
\ee
and is dominated by the top quark.  The partial widths into Goldstone
bosons are given by
\be
\Gamma_{\phi\rightarrow h h} ~=~\Gamma_{\phi\rightarrow Z_LZ_L } ~=~ \frac{1}{2}\Gamma_{\phi\rightarrow W_L^+W_L^-} = \frac{r_v^2}{32\pi}\frac{m_\phi^3}{\hat f^2}~,
\ee
where we neglect EWSB effects.  These are the dominant decay modes.

The one-loop decays into SM gauge bosons via loops of vector-like
fermions and $SO(5)/SO(4)$ composite vector bosons are given by
\be
\Gamma_{\phi\rightarrow g g}=\alpha_s^2 \, \frac{\,N^2_{\phi gg}}{72\pi^3}\frac{m_\phi^3}{\hat f^2}\,,
\ee
\be
\Gamma_{\phi\rightarrow \gamma\gamma}=\alpha^2 \, \frac{\, N_{\phi \gamma\gamma}^2}{256\pi^3}\frac{m_\phi^3}{\hat f^2}~,\quad \quad
\Gamma_{\phi\rightarrow Z_TZ_T}=\frac{\alpha^2}{s_W^4c_W^4}\, \frac{N_{\phi ZZ}^2 }{256\pi^3}\frac{m_\phi^3}{\hat f^2}~,
\ee
\be
\Gamma_{\phi\rightarrow \gamma Z_T}=\frac{ \alpha^2}{s^2_W c^2_W} \frac{N_{\gamma Z}^2 }{128\pi^3}\frac{m_\phi^3}{\hat f^2}~,\quad\quad
\Gamma_{\phi\rightarrow W_T^+W_T^-}= \frac{   \alpha^2}{s^{4}_W } \frac{N_{\phi WW}^2}{128\pi^3}\,\frac{m_\phi^3}{\hat f^2}~.
\ee
The total width for loop-induced decays into transverse electroweak
bosons can be written as
\be
\Gamma_{\phi\rightarrow V_TV'_T}= \alpha^2 \frac{\,  3 s^{-4}_W  N_{\phi WW}^2+ c^{-4}_W  N_{\phi BB}^2}{256\pi^3}\,\frac{m_\phi^3}{\hat f^2}~.
\ee

Mixed decays into one SM fermion and one of its partners may also be
possible.  The most important channels are typically the ones
involving the right handed top, and to a lesser extend the left handed
top-bottom doublet, provided the corresponding partners are not too
heavy.  Denoting the mixing angles by $s_{R}$ and $s_{L}$
respectively, one finds \footnote{ Note that the mixed Yukawa
interaction is $\mathcal L=-s_R\xi_{U,1}\phi \bar Q_L t_R+$h.c. The
vectorlike masses split into $c_R M_\psi$ (the $U$ state) and $M_\psi$
(the $Q$ state).  For large mixing, $Q$ and $U$ are approximate mass
eigenstates, and the decay proceeds to $t$ and $t'=Q$.  If the mixing
is smaller, the mass eigenstates become approximately degenerate again
and are roughly equal mixtures of $Q$ and $U$.  The interaction is
thus between $Q_L=(t'+t'')/\sqrt 2$ and $t_R$, and the decay proceeds
to two states $t'$ and $t''$.  The net effect is the same.}
\be
\Gamma_{\phi\rightarrow  t' \bar t\,,t \bar t' } =N_c\frac{|\xi_{U,1}|^2 s_{R}^2}{4\pi}\, m_\phi\,
\gamma_{\psi}^2~,
\qquad
\Gamma_{\phi\rightarrow  q' \bar q\,,q \bar q' } =N_c\frac{|\xi_{U,4}|^2 s_{L}^2}{4\pi}\, m_\phi\,
\gamma_{\psi}^2~, 
\label{eq:decay_heavy_fermion}
\ee
where we have neglected EWSB effects (in particular we neglect the top
and bottom squared masses against those of $t'$, $q'$ and $\phi$), and
we have defined $\gamma_{\psi}={1-\frac{M_\psi^2}{m_\phi^2}}$.  The
subindices $1$ and $4$ on $\xi_U$ indicate the $SO(4)$ representation
of the top partners.  They are related to the $SO(5)$ symmetric Yukawa
$\xi_U$ via the factors $w_i$ in Table {\ref{tab:yukawas4} and
\ref{tab:yukawas5}.

If the fermion resonances are sufficiently light, it is possible for
the global Higgs to decay into a heavy fermion pair.  We will give the
corresponding partial widths in a simplified limit in
Subsection~\ref{openfermioncase}.

\subsection{Case I: Closed  Decay Channels into  Fermion Resonances}

%
\begin{figure}
\centering
\begin{picture}(400,120)
\put(-20,0){
\put(0,0){\includegraphics[scale=0.6,clip=true, trim= 0cm 0cm 0cm 0cm]{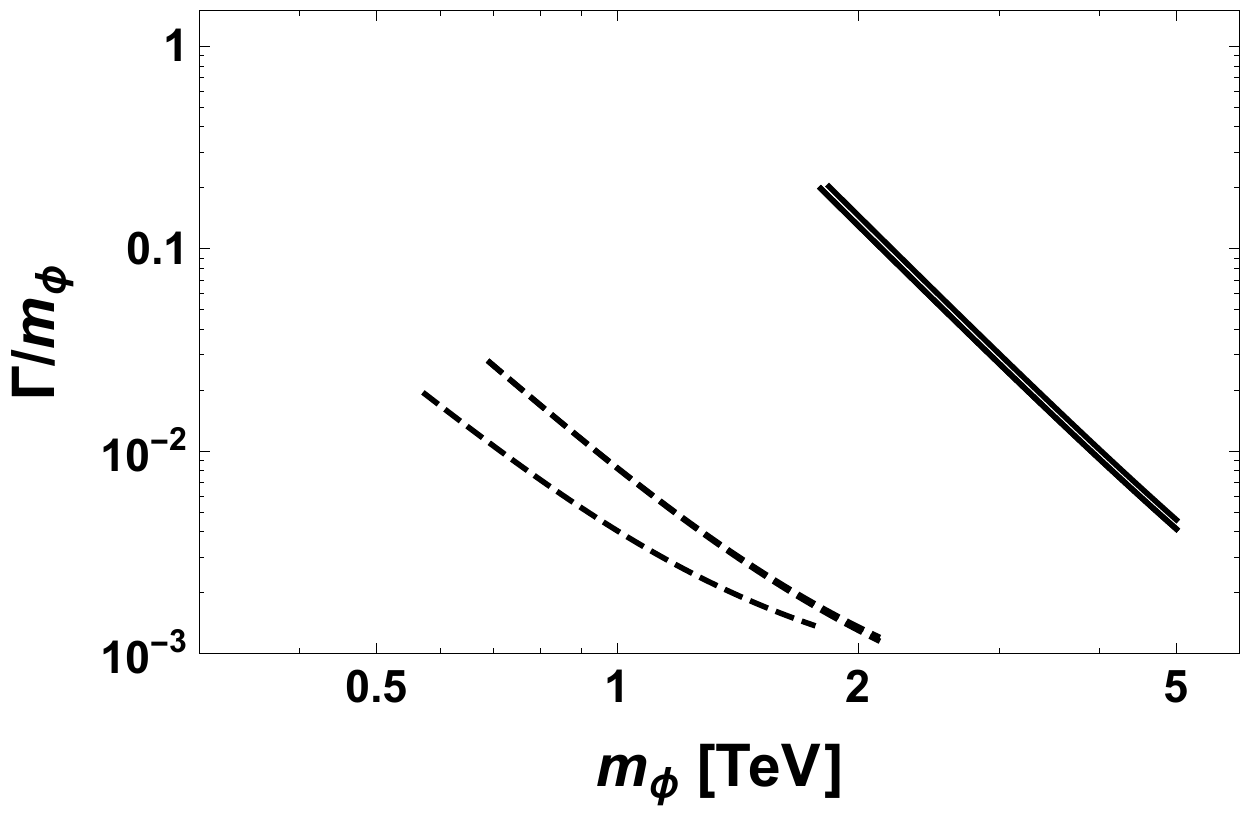}\quad
\includegraphics[scale=0.6,clip=true, trim= 0cm 0cm 0cm 0cm]{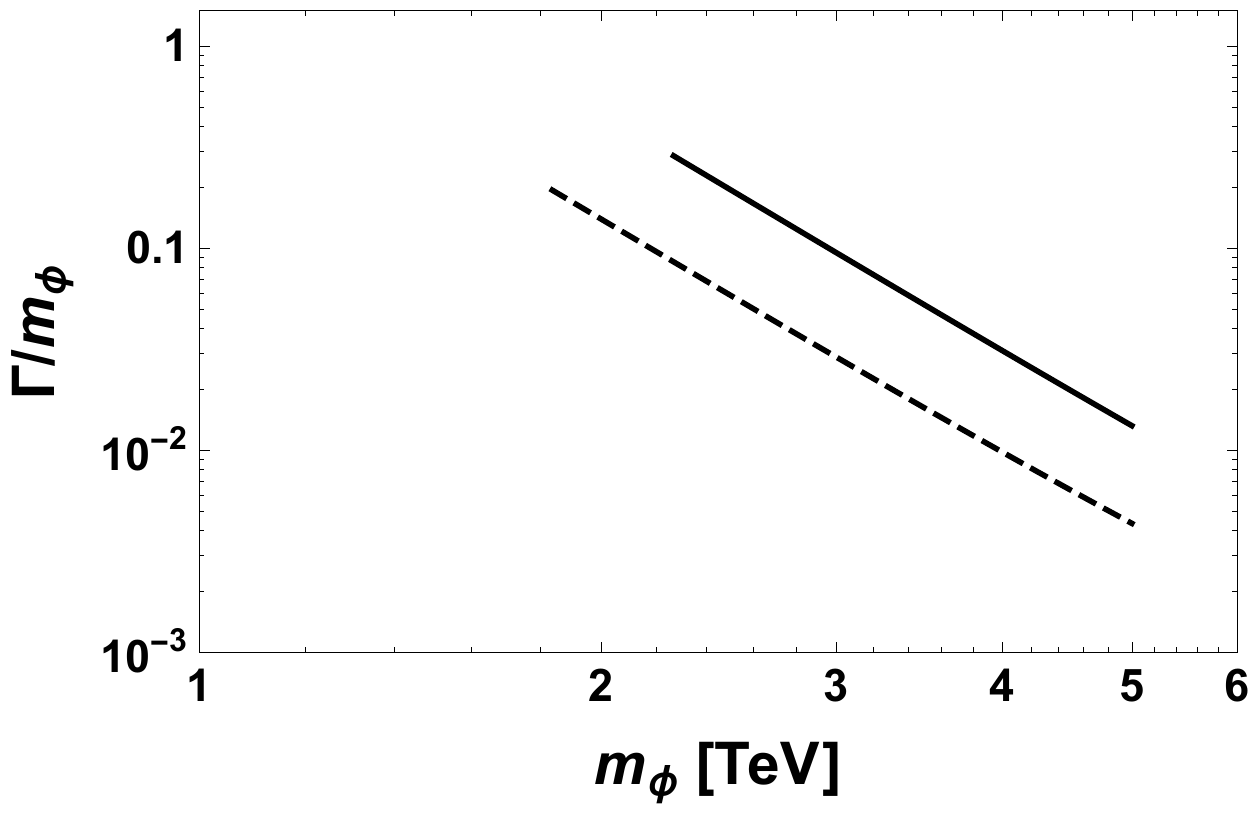}}
\put(70,93){{\color{red} \boxed{$ \Scale[0.8]{\lambda = \xi^2}$}}}
\put(160,120){{\color{red} \boxed{$ \Scale[0.8]{\lambda_{\rm max}}$}}}
\put(125,65){{\color{orange} $ \Scale[0.6]{(5,1,10)},$}}
\put(120,55){{\color{orange} $ \Scale[0.6]{(14,14,10)}$}}
\put(65,50){{\color{orange} $ \Scale[0.6]{(5,14,10)}$}}
\put(180,95){{\color{orange} $ \Scale[0.6]{(5,1,10)},$}}
\put(180,85){{\color{orange} $ \Scale[0.6]{(5,14,10)},$}}
\put(180,75){{\color{orange} $ \Scale[0.6]{(14,14,10)}$}}

\put(295,120){{\color{red} \boxed{$ \Scale[0.8]{\lambda = \xi^2}$}}}
\put(360,120){{\color{red} \boxed{$ \Scale[0.8]{\lambda_{\rm max}}$}}}
\put(340,80){{\color{orange} $ \Scale[0.6]{(5,1)}$}}
\put(400,85){{\color{orange} $ \Scale[0.6]{(5,1)}$}}

}
\end{picture}
\caption{Total width of the global Higgs in the case that the decays
to fermion resonances are forbidden.  The plot shows curves for
$\lambda=\xi^2$ (dashed lines) and $\lambda = \lambda_{\rm max}$
(continuous lines), as given in Table~\ref{tab:couplings} for the
different models.  The minimum $m_\phi$ is determined by $\hat f \geq
f$.
\label{fig:Gammatot}
}
\end{figure}
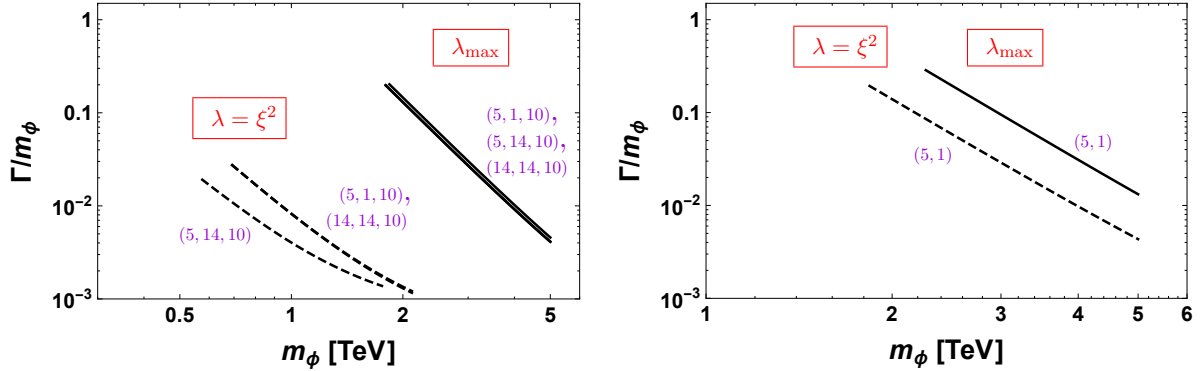
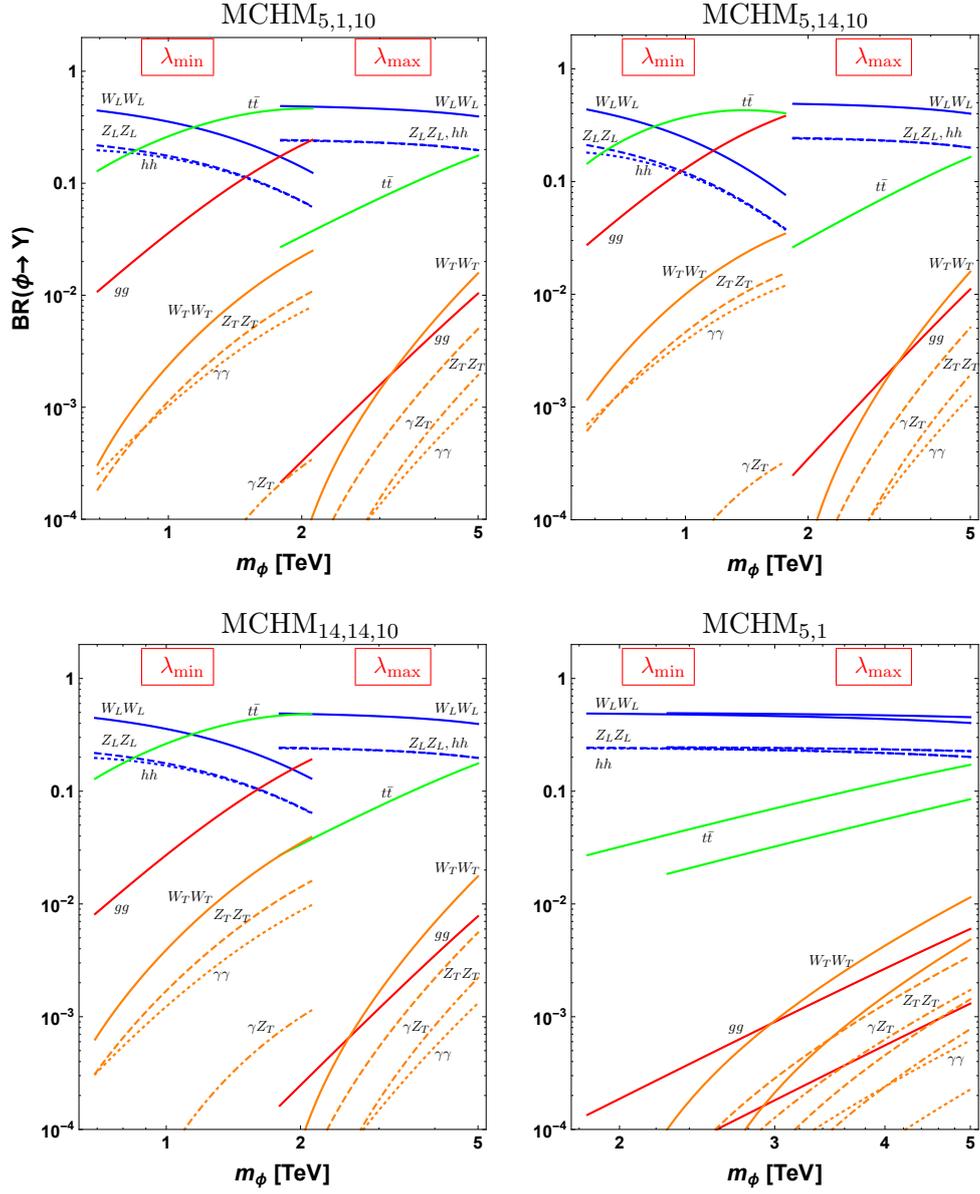
\begin{figure}[t]
\begin{center}
\begin{picture}(500,420)
\put(30,0){
\put(0,230){\includegraphics[scale=0.5,clip=true, trim= 0cm 0cm 0cm 0cm]{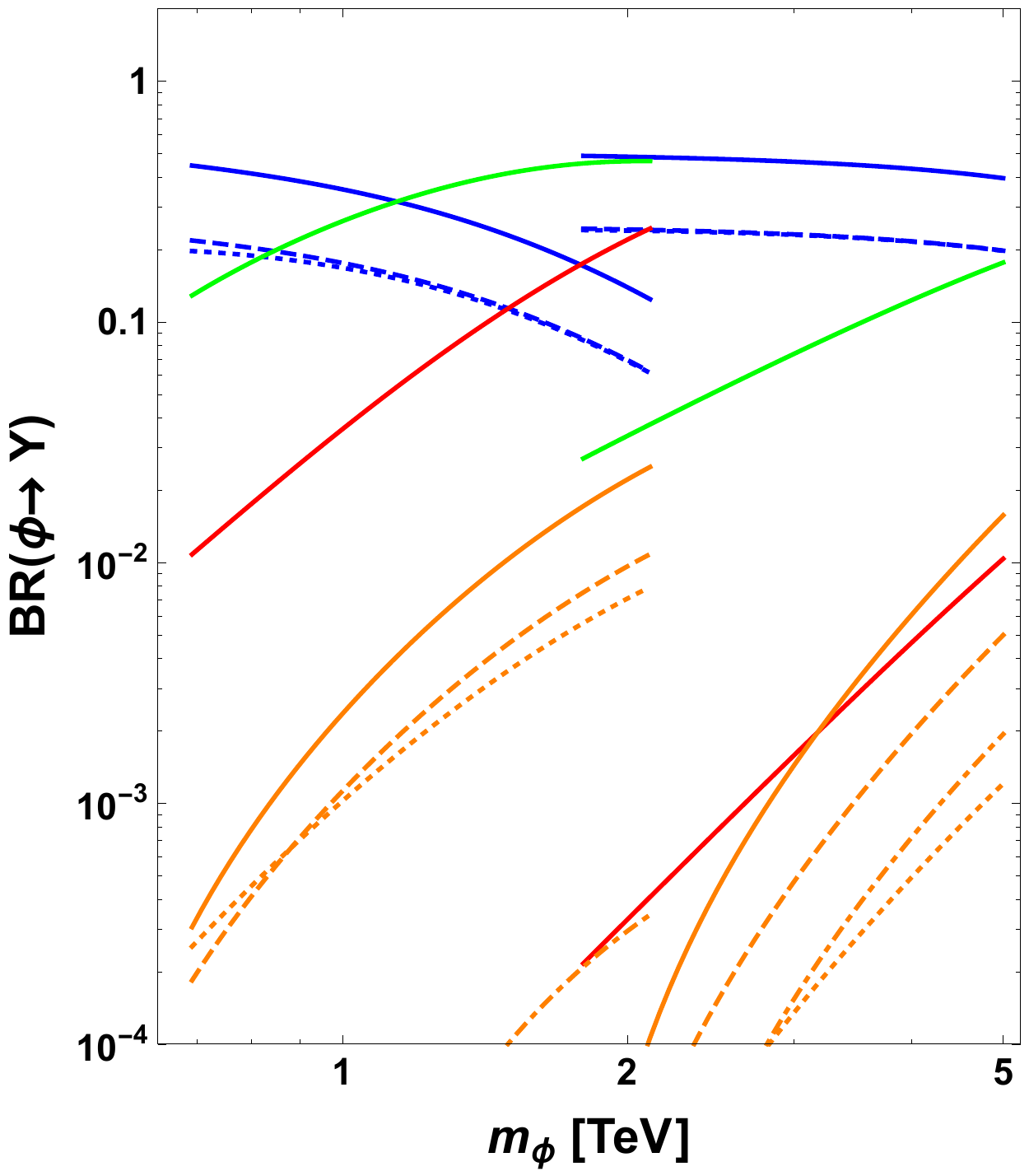}
\includegraphics[scale=0.5,clip=true, trim= 0cm 0cm 0cm 0cm]{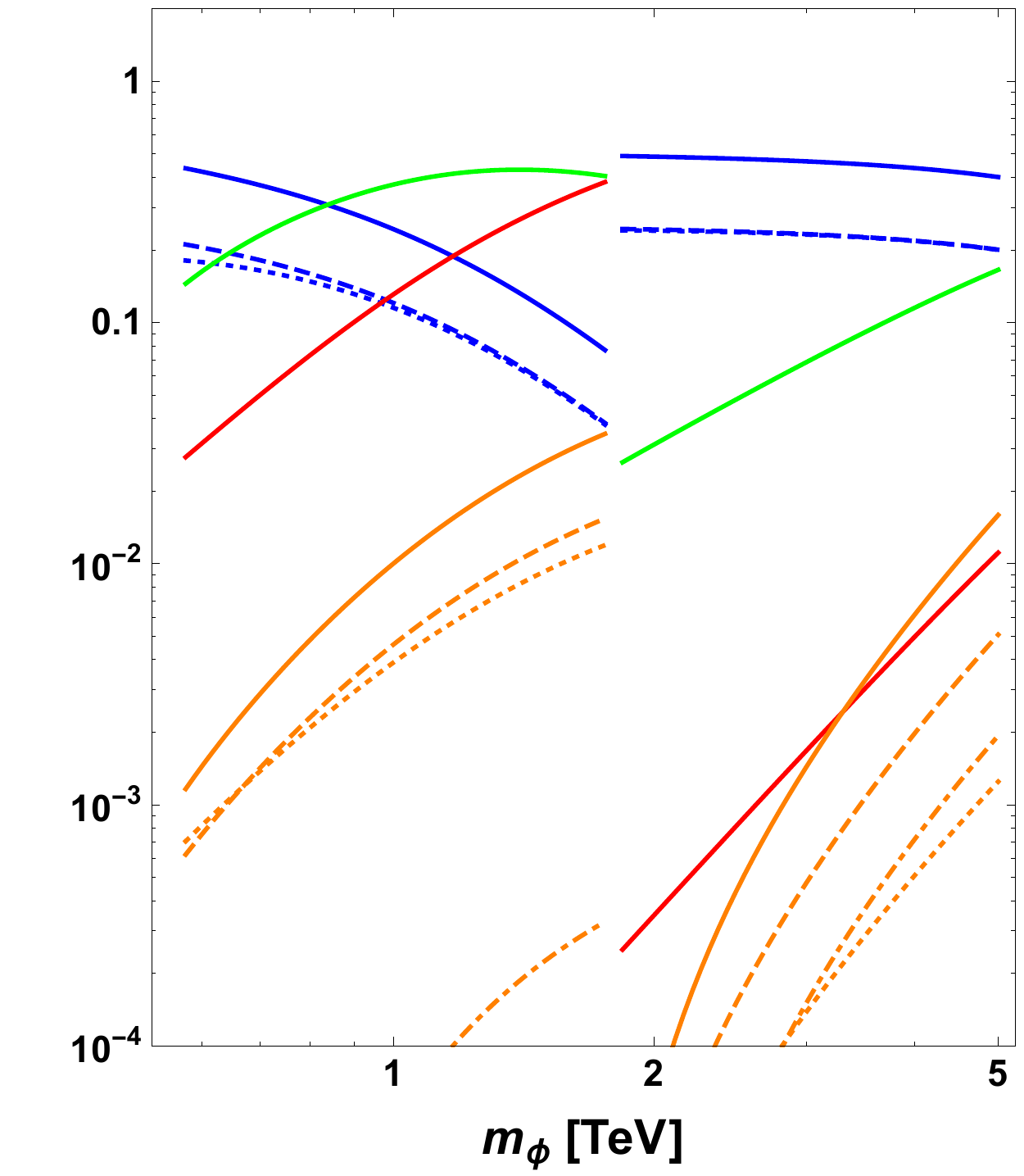}}  
\put(0,0){\includegraphics[scale=0.5,clip=true, trim= 0cm 0cm 0cm 0cm]{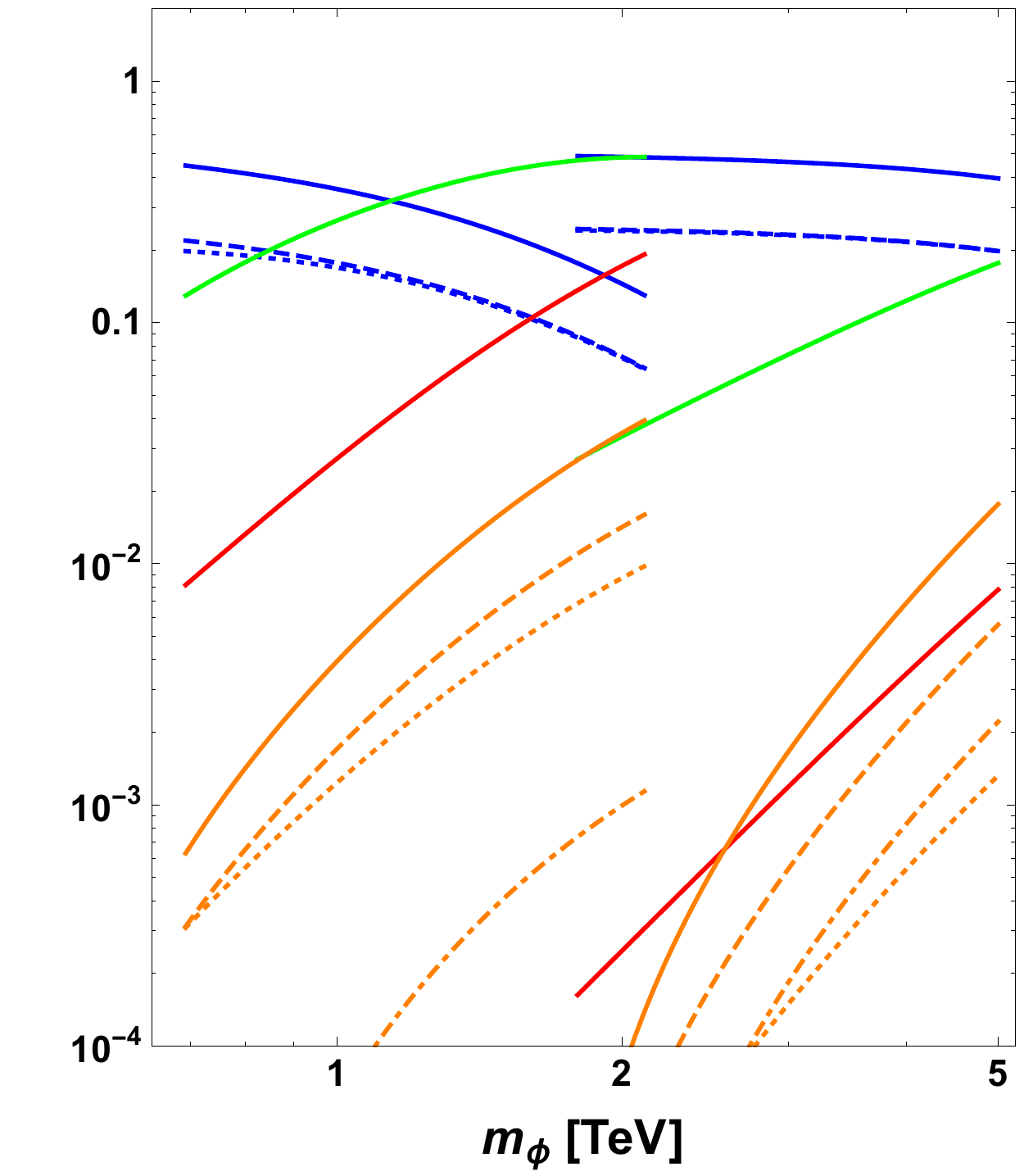}
\includegraphics[scale=0.5,clip=true, trim= 0cm 0cm 0cm 0cm]{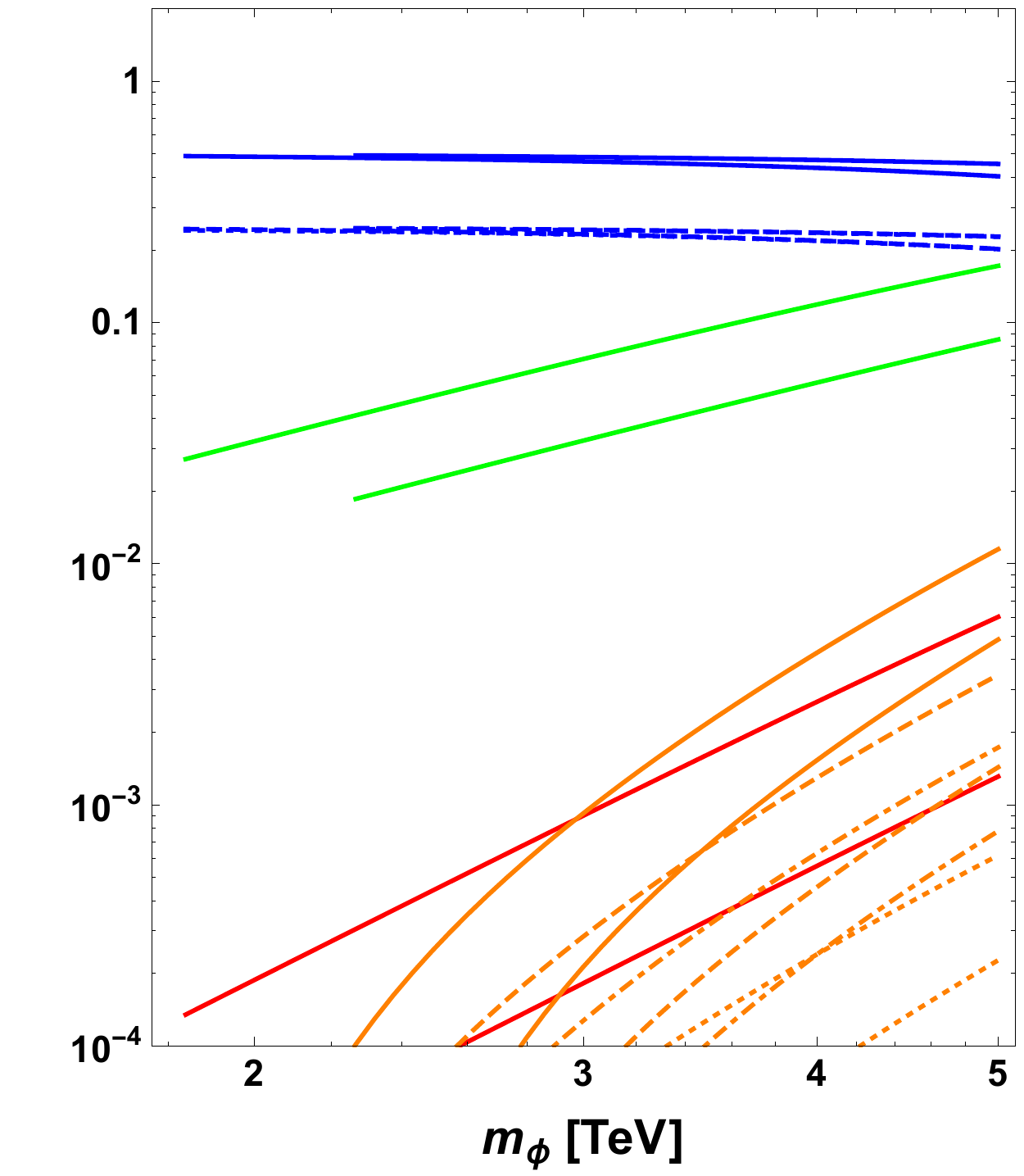}
}

\put(80,440){{\color{black} MCHM$_{5,1,10}$}}
\put(260,440){{\color{black} MCHM$_{5,14,10}$}}
\put(80,210){{\color{black} MCHM$_{14,14,10}$}}
\put(260,210){{\color{black} MCHM$_{5,1}$}}

\put(50,195){{\color{red} \boxed{$ \Scale[0.8]{\lambda_{\rm min}}$}}}
\put(130,195){{\color{red} \boxed{$ \Scale[0.8]{\lambda_{\rm max}}$}}}
\put(230,195){{\color{red} \boxed{$ \Scale[0.8]{\lambda_{\rm min}}$}}}
\put(310,195){{\color{red} \boxed{$ \Scale[0.8]{\lambda_{\rm max}}$}}}

\put(50,425){{\color{red} \boxed{$ \Scale[0.8]{\lambda_{\rm min}}$}}}
\put(130,425){{\color{red} \boxed{$ \Scale[0.8]{\lambda_{\rm max}}$}}}
\put(230,425){{\color{red} \boxed{$ \Scale[0.8]{\lambda_{\rm min}}$}}}
\put(310,425){{\color{red} \boxed{$ \Scale[0.8]{\lambda_{\rm max}}$}}}

\put(35,180){{\color{black} $ \Scale[0.5]{W_LW_L}$}}
\put(35,168){{\color{black} $ \Scale[0.5]{Z_LZ_L}$}}
\put(50,155){{\color{black} $ \Scale[0.5]{hh}$}}
\put(90,179){{\color{black} $ \Scale[0.5]{t\bar t}$}}
\put(40,105){{\color{black} $ \Scale[0.5]{gg}$}}
\put(77,80){{\color{black} $ \Scale[0.5]{\gamma\gamma}$}}
\put(90,60){{\color{black} $ \Scale[0.5]{\gamma Z_T }$}}
\put(77,102){{\color{black} $ \Scale[0.5]{Z_T Z_T}$}}
\put(60,109){{\color{black} $ \Scale[0.5]{W_T W_T}$}}

\put(160,180){{\color{black} $ \Scale[0.5]{W_LW_L}$}}
\put(150,167){{\color{black} $ \Scale[0.5]{Z_LZ_L, hh}$}}
\put(140,148){{\color{black} $ \Scale[0.5]{t\bar t}$}}
\put(160,95){{\color{black} $ \Scale[0.5]{gg}$}}
\put(160,50){{\color{black} $ \Scale[0.5]{\gamma\gamma}$}}
\put(148,60){{\color{black} $ \Scale[0.5]{ \gamma Z_T }$}}
\put(163,80){{\color{black} $ \Scale[0.5]{Z_T Z_T}$}}
\put(160,120){{\color{black} $ \Scale[0.5]{W_T W_T}$}}

\put(35,410){{\color{black} $ \Scale[0.5]{W_LW_L}$}}
\put(35,398){{\color{black} $ \Scale[0.5]{Z_LZ_L}$}}
\put(50,385){{\color{black} $ \Scale[0.5]{hh}$}}
\put(90,408){{\color{black} $ \Scale[0.5]{t\bar t}$}}
\put(40,338){{\color{black} $ \Scale[0.5]{gg}$}}
\put(77,307){{\color{black} $ \Scale[0.5]{\gamma\gamma}$}}
\put(90,265){{\color{black} $ \Scale[0.5]{\gamma Z_T }$}}
\put(80,326){{\color{black} $ \Scale[0.5]{Z_T Z_T}$}}
\put(60,330){{\color{black} $ \Scale[0.5]{W_T W_T}$}}

\put(160,409){{\color{black} $ \Scale[0.5]{W_LW_L}$}}
\put(150,397){{\color{black} $ \Scale[0.5]{Z_LZ_L, hh}$}}
\put(140,378){{\color{black} $ \Scale[0.5]{t\bar t}$}}
\put(160,320){{\color{black} $ \Scale[0.5]{gg}$}}
\put(160,277){{\color{black} $ \Scale[0.5]{\gamma\gamma}$}}
\put(149,288){{\color{black} $ \Scale[0.5]{ \gamma Z_T }$}}
\put(165,310){{\color{black} $ \Scale[0.5]{Z_T Z_T}$}}
\put(160,348){{\color{black} $ \Scale[0.5]{W_T W_T}$}}

\put(220,409){{\color{black} $ \Scale[0.5]{W_LW_L}$}}
\put(215,396){{\color{black} $ \Scale[0.5]{Z_LZ_L}$}}
\put(235,383){{\color{black} $ \Scale[0.5]{hh}$}}
\put(275,409){{\color{black} $ \Scale[0.5]{t\bar t}$}}
\put(225,358){{\color{black} $ \Scale[0.5]{gg}$}}
\put(262,322){{\color{black} $ \Scale[0.5]{\gamma\gamma}$}}
\put(275,272){{\color{black} $ \Scale[0.5]{\gamma Z_T }$}}
\put(265,341){{\color{black} $ \Scale[0.5]{Z_T Z_T}$}}
\put(245,345){{\color{black} $ \Scale[0.5]{W_T W_T}$}}

\put(345,409){{\color{black} $ \Scale[0.5]{W_LW_L}$}}
\put(335,397){{\color{black} $ \Scale[0.5]{Z_LZ_L, hh}$}}
\put(325,377){{\color{black} $ \Scale[0.5]{t\bar t}$}}
\put(345,320){{\color{black} $ \Scale[0.5]{gg}$}}
\put(345,277){{\color{black} $ \Scale[0.5]{\gamma\gamma}$}}
\put(334,288){{\color{black} $ \Scale[0.5]{ \gamma Z_T }$}}
\put(350,310){{\color{black} $ \Scale[0.5]{Z_T Z_T}$}}
\put(345,348){{\color{black} $ \Scale[0.5]{W_T W_T}$}}

\put(220,182){{\color{black} $ \Scale[0.5]{W_LW_L}$}}
\put(220,170){{\color{black} $ \Scale[0.5]{Z_LZ_L}$}}
\put(220,159){{\color{black} $ \Scale[0.5]{hh}$}}
\put(260,131){{\color{black} $ \Scale[0.5]{t\bar t}$}}
\put(270,60){{\color{black} $ \Scale[0.5]{gg}$}}
\put(352,48){{\color{black} $ \Scale[0.5]{\gamma\gamma}$}}
\put(322,60){{\color{black} $ \Scale[0.5]{\gamma Z_T }$}}
\put(335,70){{\color{black} $ \Scale[0.5]{Z_T Z_T}$}}
\put(300,85){{\color{black} $ \Scale[0.5]{W_T W_T}$}}

}
\end{picture}
\caption{Branching fractions of the global Higgs in the
MCHM$_{5,1,10}$, MCHM$_{5,15,10}$, MCHM$_{14,14,10}$, MCHM$_{5.1}$
scenarios, assuming that decays into fermion resonances are forbidden.
Both extreme values $\lambda=\{\xi^2,\lambda_{\rm max}\}$ of the
global Higgs quartic coupling are shown, and we fix $f = 800~{\rm
GeV}$ and $M_\psi = m_\phi$.  Blue lines correspond to $W_LW_L$
(solid), $Z_LZ_L$ (dashed), $hh$ (dotted) final states.  The green
line is $t\bar t$.  The red line is $gg$ and orange lines correspond
to $W_TW_T$ (solid), $Z_TZ_T$ (dashed), $\gamma\gamma$ (dotted),
$\gamma Z_T$ (dash-dotted).  The minimum $m_\phi$ is determined by
$\hat f \geq f$.
\label{fig:BRs}}
\end{center}
\end{figure}

We assume first that the decays of the global Higgs into SM fermion
partners are kinematically forbidden, e.g.~$M_i>m_\phi$.  This
assumption also has implications for the loop decays, which are
controlled by the relative contribution to the fermion masses from
global symmetry breaking versus symmetry preserving effects, as
described in Sec.~\ref{se:oneloop}.  This relative importance is
characterized by $(\xi \hat{f})^2/M^2_\psi = (\xi^2/2\lambda)
(m_\phi/M_\psi)^2 \lesssim \xi^2/2\lambda$, which can be seen to be at
most of order one in the lower end of the range for $\lambda$ (see
Table~\ref{tab:couplings}), hence the approximate formulas
Eqs.~\eqref{eq:coupgg}, \eqref{eq:coupgamgam} are valid.  In most of
the considered range for $\lambda$, the $(\xi \hat{f})^2/M^2_\psi$
factor will in fact induce an important suppression for such decays,
in addition to the 1-loop suppression.  For illustration, we will use
$M_\psi = m_\phi$ in Eqs.~(\ref{eq:coupgg}) and (\ref{eq:coupgamgam}),
and take the value $A_{1/2}(1/4) \approx 1.42$ for the fermion loop
function (slightly above the asymptotic value of 4/3).

We then have a rather predictive case, since the dominant features
depend on only three parameters, that can be taken as the global Higgs
mass $m_\phi$, the quartic coupling $\lambda$, and the Higgs decay
constant $f$.  The latter controls the deviations of the pNGB Higgs
properties from the SM limit, and can be constrained by Higgs 
measurements, which as illustrated in~\cite{Carena:2014ria}, can be 
fairly model-dependent. For concreteness, we will take $f = 800~{\rm GeV}$, 
which should allow to satisfy comfortably the current Higgs constraints for 
a wide choice of parameters in the fermionic sector. In addition, such a 
choice also allows for generic consistency with EW precision measurements 
(see, for example,~\cite{Azatov:2013ura}). A more detailed study of Higgs 
and EW precision constraints is beyond the scope of this work, and is not 
expected to change our conclusions. Thus, fixing $f$ allows us to focus on 
the properties of the global Higgs, as controlled by the two remaining parameters, 
$m_\phi$ and $\lambda$, which barely affect the SM Higgs phenomenology.\footnote{
The low-energy effects of the global Higgs are described by
loop-generated dimension-6 operators and tree-level dimension-8
operators.} Note that Eqs.~(\ref{eq:GH_mass}) and (\ref{rv}) imply
that $m_\phi \geq \sqrt{2\lambda} f$, so that for a given value of
$\lambda$ one obtains a minimum global Higgs mass. One could also be worried about 
potential direct lower limits on $m_\phi$. Adapting the ATLAS heavy Higgs search of
Ref.~\cite{ATLAS-CONF-2013-013}, we obtain  that the global Higgs
must be roughly heavier than about $750$~GeV \cite{wip}.

An interesting feature of the global Higgs couplings to transverse
electroweak gauge bosons is that they are dominated by the loops of
the spin-1 (coset) resonances for large values of $\lambda$.  In this
case these couplings are mainly controlled by the $r_v$ parameter (the
spin-1 amplitude scales as $1-r_v$, as can be seen from
Eq.~\eqref{Nphigaga}), up to small corrections from the fermion loops,
and thus depend only mildly on the fermion sector.  On the other hand
the EW couplings at small $\lambda$, as well as the gluon coupling,
are fully dependent on the sector of fermion resonances.

The total width of the global Higgs is dominated by the decays into
the $SO(5)/SO(4)$ Goldstone bosons and into pairs of top quarks.
These contributions do not depend on the details of the fermion
sector, so that one has in general
\bea
\frac{\Gamma_{\rm tot}}{m_{\phi}} &\approx& \frac{r_v^2}{32\pi} \frac{m_\phi^2}{\hat f ^2} +\frac{3\,m_t^2}{8\pi \hat f ^2}~,
\eea
to a very good approximation.  The total widths are shown in
Fig.~\ref{fig:Gammatot}.  We use the relation $m_\phi^2/{\hat
f}^2=2\lambda$, together with the estimates of $\lambda$ derived in
Sec.~\ref{se:perturbativity}.  It turns out that the total width
ranges from $\Gamma_\phi/m_\phi = {\cal O} (10^{-3})$ to ${\cal
O}(0.1)$.  The global Higgs is thus always narrow enough so that the
``narrow width approximation'' applies.

We also show in Fig.~\ref{fig:BRs} the branching fractions for our
benchmark scenarios, displayed as a function of $m_\phi$.  We do this
for the two extreme estimates of the global Higgs quartic coupling
$\lambda = \xi^2$ and $\lambda = \lambda_{\rm max}$, as determined in
Sec.~\ref{se:perturbativity}.  We observe that
$\Gamma_{\phi\rightarrow \gamma\gamma}$ is smaller than
$\Gamma_{\phi\rightarrow W^+W^-}$ and $\Gamma_{\phi\rightarrow ZZ}$ by
several orders of magnitude at $m_\phi=750$ GeV. Due to the LHC13
bounds on the diboson $ZZ$, $WW$ channels from
ATLAS~\cite{Aaboud:2016okv} and CMS \cite{CMS:2015nmz}, the
possibility of interpreting the $750$~GeV diphoton excess
\cite{atlasdiphoton,CMS:2015dxe} as originating from the resonant
production of a narrow global Higgs with $m_\phi=750$~GeV is excluded.
It turns out that the branching fractions into two gluons, two
photons, and into the transverse components of the weak gauge bosons
become more important for a heavier global Higgs (see
Fig.~\ref{fig:BRs}).  Such enhancement of the couplings to transverse
gauge bosons is potentially interesting for production of the global
Higgs at the LHC and will be explored in more detail in the
accompanying Ref.~\cite{wip}.

\subsection{Case II: Open  Decay Channels into  Fermion Resonances}
\label{openfermioncase}

Clearly, when decays into fermionic resonances (or mixed decays into a
SM fermion and one of its partners) are kinematically open, the
branching fractions are sensitive to the details of the new fermionic
sector.  For illustration, we consider the case where all the fermion
resonances are light compared to the global Higgs.  In this case, all
possible two-body decay channels are open.  Neglecting the small
$M_\psi^2/m_\phi^2$ terms, \ie~taking all $\gamma_\psi=1$ in
Eq.~(\ref{eq:decay_heavy_fermion}), and assuming universal $SO(5)$
proto-Yukawa couplings $\xi=\xi_{U,D}=\xi'_{U,D}$, the fermion mixing
angles appearing in Eq.~\eqref{eq:decay_heavy_fermion} simplify.  The
decays into heavy fermion pairs then contribute to the total width as
\bea
\frac{\Gamma_{\phi\rightarrow \psi \bar \psi}}{m_{\phi}} &=&
\frac{27}{4\pi} |\xi|^2\approx 0.8
 \quad  ({\rm MCHM}_{5,1,10} )~,
\\
\frac{\Gamma_{\phi\rightarrow \psi \bar \psi}}{m_{\phi}} &=&
\frac{54}{5\pi} |\xi|^2\approx 0.9
 \quad  ({\rm MCHM}_{5,14,10} )~,
\\
\frac{\Gamma_{\phi\rightarrow \psi \bar \psi}}{m_{\phi}} &=&
\frac{117}{20\pi}|\xi|^2 \approx 0.7
 \quad  ({\rm MCHM}_{14,14,10} )~,
\\
\frac{\Gamma_{\phi\rightarrow \psi \bar \psi}}{m_{\phi}} &=&
\frac{3}{4\pi} |\xi|^2\approx 0.6
 \quad ({\rm MCHM}_{5,1} )~,
\eea
where $\xi$ (at $\mu=m_\phi$) has been estimated in
Sec.~\ref{se:perturbativity} for each benchmark scenario.  We conclude
that when several fermionic decays are open, the global Higgs is in
general a broad resonance, unless all such decays occur very near
threshold and there is a further kinematic suppression.

\section{Conclusions} 
\label{se:conclusions}

We investigated the properties of the physical excitations of the
global symmetry breaking vacuum in composite Higgs models.  Such a
\textit{global Higgs} is expected to interact with the SM Higgs and
electroweak gauge bosons, with the SM fermions proportionally to their
mass, and with the heavy fermion and vector resonances of the theory.
An effective coupling to photons, gluons and transverse electroweak
gauge bosons via loops of the resonances is also expected.

We studied in detail the minimal $SO(5)/SO(4)$ case through a general
2-sites model Lagrangian, and found that the dominant interactions of
the global Higgs with the SM particles are controlled by two
real-valued parameters and by a few group theoretical factors.  The
couplings of the global Higgs to the SM fermions depend on the global
Higgs decay constant $\hat f$ and on whether the proto-Yukawa
structure is linear or bilinear in $\Phi$, the $SO(5)$ multiplet
containing the global Higgs.  The couplings of the global Higgs to the
pNGBs depend on $\hat f$, on the usual NGB decay constant $f$, and on
the global symmetry group.  In a large region of parameter space, the
dominant decay modes of the global Higgs are the tree-level decays to
the SM Higgs, electroweak gauge bosons, and top quark.

The global Higgs also couples to the (possibly many) fermion
resonances that partner with the SM fermions.  We analyzed various
typical realizations of the $SO(5)$ fermionic sector, with a global
Higgs arising either from the $\bf 5$ or $\bf 14$ of $SO(5)$.  We
computed the beta functions of the composite sector, \ie~the global
Higgs quartic and the $SO(5)$ Yukawa couplings.  Evolving these
couplings from the strong coupling scale down to the global Higgs mass
scale provides a consistent picture of the composite sector, necessary
for the analysis of the global Higgs properties.

Loops of fermion and vector resonances of the coset induce an
effective coupling of the global Higgs to SM gauge bosons.
This is similar to the case of the Higgs-photon coupling induced by
top quark and $W$ loops, except that for the global Higgs the fermion
multiplicity can be much larger, enhancing the loop amplitude
accordingly.  We derived compact formulas for these effective couplings
in each realization of the fermion sector in the benchmark models
considered.

When several heavy fermion channels are open, the global Higgs is in
general a broad resonance.  On the other hand, when the decay of the
global Higgs into fermion resonances is kinematically suppressed or
forbidden, its decay width ranges from $\Gamma_{\rm tot}/m_\phi\sim
10^{-3}$ to $\sim 0.1$, depending on the global Higgs mass and quartic
coupling.  The global Higgs can thus behave either as a narrow or a
broad resonance.  In this latter more predictive case, we provided the
branching fractions of the global Higgs for each benchmark model.

Although the present study is mostly theoretical, it turns out that
the properties of the global Higgs are such that it could in principle
be detected at a collider like the LHC. That is, the theoretical
aspects of composite Higgs models we explored here may turn into a new
way of searching for Higgs compositeness at the LHC. A detailed study
of the collider implications of a global Higgs is presented in
Ref~\cite{wip}.  As a motivation, we simply observe that the coupling
of the global Higgs to gluons, induced by the many fermion resonances
of the theory, may be sizeable enough to allow for the production of
the global Higgs by gluon fusion at the LHC with $300$~fb$^{-1}$
integrated luminosity.

\acknowledgments 
This work was supported by the S\~ao Paulo Research Foundation
(FAPESP) under grants \#2011/11973 and \#2014/21477-2.  E.P. and
R.R.~were partially funded by a CNPq research grant.

\appendix

\section{The Global Higgs in the $\bf 14$ Representation of $SO(5)$}
\label{se:14}

In Sec.~\ref{se:boson_couplings} and below, we have assumed that the
global Higgs is embedded in a fundamental $\Phi={\bf 5}$ of $SO(5)$,
\ie~it is identified with the $SO(4)$ singlet in the decomposition
\be
\bf 5\to (2,2)+(1,1)\,.
\ee
We parametrized this decomposition by the NGB matrix $U_5$ and the
radial direction $\H$ as
\be
\Phi=U_5 \H\,,
\ee
and aligned $\H$ as 
\be
\H=(\hat f+\phi)e_5\,,\qquad e_5=(0,0,0,0,1)^T\,.
\ee
This is the most minimal scenario possible.

The next-to-minimal embedding is in the symmetric traceless {$\Psi=\bf
14$} representation.  Indeed, the decomposition into $SO(4)$ also
contains an $SO(4)$ singlet:
\be
\bf 14\to (3,3)+(2,2)+(1,1)
\ee
which we will be parametrizing as follows:\footnote{Unlike the vacuum
induced by a vev in the {\bf 5}, which is unique, the breaking by the
$\bf 14$ can also lead to other vacua such as $SO(3)\times SO(2)$.  We
assume here that there exists a potential that leads to the $SO(4)$
vacuum.}
\be
\Psi=U_5\, (\H+\H')\, U_5^\dagger~.
\ee
Here $\H'$ denotes the $(\bf 3,3)$
\be
\H'=\left(\begin{matrix}
\phi'_{4\times 4}&\\&0
\end{matrix}\right)~,
\ee
with $\phi'$ traceless symmetric, and leads to (non-NGB) heavy states.
The singlet is parametrized as
\be
\H=(\hat f+ \phi)e_{14}\,,\qquad
e_{14}=
\begin{pmatrix}
\frac{1}{2\sqrt{5}}\times 1_{4\times 4} &\\
&-\frac{2}{\sqrt{5}}
\end{pmatrix}~.
\ee
Notice that $\Tr e_{14}=0$ and $\Tr e_{14}^2=1$.

A first comment regards the scalar potential.  There are now two
independent quartic couplings that are conveniently written as
\be
V=\frac{\lambda}{4}\left(\tr \Psi^2-\hat f^2\right)^2+\frac{\lambda'}{4}\left(\frac{13}{5}[\tr \Psi^2]^2-4\tr \Psi^4\right)~.
\ee
This potential contains an $SO(4)$ symmetric vacuum
$\langle\phi'\rangle=0$ for $\lambda'>0$, $\lambda>0$ with
\be
m_1^2=2\lambda\hat f^2~,\qquad m_9^2=2\lambda' \hat f^2~.
\ee
We will assume that $\lambda'$ is sufficiently large so as to decouple
the nonet near the cutoff.

A second modification concerns the vector resonances.
Eq.~(\ref{Lbos}) is then still valid provided we use the corresponding
covariant derivative
\be
\nabla \H=\partial_\mu\H-i\mathcal A_\mu^{\hat a} \left[T^{\hat a}\,,\,\H\right]~,
\ee
such that Eq.~(\ref{eq:GBkin}) gets modified according to
\begin{multline}
\frac{1}{2}|\nabla \H|^2+\frac{1}{4}f_\rho^2
\left( \A^A_\mu-i[ U_{5}^\dagger  D_\mu U_{ 5}]^A \right)^2\\
=\frac{1}{2}(\partial_\mu\phi)^2+\frac{5}{8}(\hat f+\phi)^2(\A_\mu^{\hat a})^2+\frac{f_\rho^2}{4}\left(\A^{\hat a}_\mu+\frac{\sqrt{2}}{f}D_\mu h^{\hat a}\right)^2+\cdots
\end{multline}
Proceeding similarly to Eq.~(\ref{eq:GBkin}) one obtains
\be
\mathcal L=
\frac{1}{2} (\partial_\mu\phi)^2+\frac{1}{2}(D_\mu h^{\hat a})^2+\frac{f_\rho^2\hat f^2}{4Zf^2}(\mathcal B_\mu^{\hat a})^2+\frac{1}{Z}\left(\frac{1}{2}\hat f \phi+\frac{1}{4}\phi^2\right)
\left(\mathcal B_\mu^{\hat a}-\frac{\sqrt{2}Z\,f}{\hat f^2}D_\mu h^{\hat a}\right)^2~,
\ee
with $Z=\frac{2}{5}$ and
\be
f^{-2}= Z \hat f^{-2}+f_\rho^{-2}\,.
\ee
We will define 
\be 
r_v\equiv\frac{m_\rho^2}{m_a^2} =\frac{Z f^2}{\hat f^2}\leq 1~,
\ee
yielding 
\be
\mathcal L=
\frac{1}{2} (\partial_\mu\phi)^2+\frac{1}{2}(D_\mu h^{\hat a})^2+\frac{m_a^2}{2g_\rho^2}(\mathcal B_\mu^{\hat a})^2+\left(\frac{\phi}{\hat f}+\frac{1}{2}\frac{\phi^2}{\hat f^2}\right)
\left(\frac{\sqrt{m_a^2-m_\rho^2}}{g_\rho}\mathcal B_\mu^{\hat a}-\sqrt{r_v}\,D_\mu h^{\hat a}\right)^2~,
\ee
as in the case of the $\bf 5$ representation.

\section{Yukawa Structures}
\label{app:Yukawas}

In this appendix we setup the conventions necessary to derive the
individual Yukawa couplings, in particular the weights in
Tables~\ref{tab:yukawas4} and \ref{tab:yukawas5}.  The $SO(5)$ fields
are parametrized as ($i=1...4$, $j=1...6$, $k=1...9$)
\bea
F&=& f_i\ e_{(4)}^i+ s\ e_{(1)}~, \nn\\
A&=& a_j\ \alpha^j_{(6)}+ f_i\  \alpha^i_{(4)}~,\\
B&=& b_k\ \beta_{(9)}^k +  f_i\  \beta^i_{(4)} + s\ \beta_{(1)}~,\nn
\eea
where the $e$'s are unit vectors, and the $\alpha$'s ($\beta$'s) are
antisymmetric (symmetric traceless) orthogonal matrices that are
normalized as $\tr \alpha^2=1$ and $\tr \beta^2=1$.  Here we only need
the explicit forms of:
\bea
(\beta_{(1)})_{ab}&=&\frac{1}{2\sqrt{5}}\delta_{ab}-\frac{2}{\sqrt{5}}\delta_{a5}\delta_{b5}~,\nonumber\\
(\alpha_{(4)}^i)_{ab}&=&\frac{1}{\sqrt2}(\delta_{ia}\delta_{b5}-\delta_{a5}\delta_{ib})~,\\
(\beta_{(4)}^i)_{ab}&=&\frac{1}{\sqrt2}(\delta_{ia}\delta_{b5}+\delta_{a5}\delta_{ib})~.\nonumber
\eea
With these conventions, kinetic terms are canonically normalized when
written as traces.  The $SO(5)$ Yukawa couplings are normalized as
\be
\mathcal L=-\xi\, \Phi_i \left( \bar F_i\,P_R\, S+ \bar F_j \,P_R\,B_{ij} + \bar F_{j}\,P_R\,A_{ij}\right)
-\xi\,\Psi_{ij}\left( B_{jk}\,P_R\, B'_{ki}+ B_{jk}\,P_R\, A_{ki}\right) +h.c.
\ee
The $SO(4)$ Yukawas are normalized as
\be
\mathcal L=-\xi\,\phi \left (\bar b_k\,P_R\, b_k'+\bar a_j \,P_R\, a_j'+ \bar f_i \,P_R\, f_i'+ \bar s\,P_R\,  s' \right)+h.c.
\ee
By comparison, one obtains the weights given in Table
\ref{tab:yukawas4} and \ref{tab:yukawas5}.

\section{Loop Functions}
\label{loopfunctions}

For completeness, we collect here the well-known loop functions (see
\cite{Djouadi:2005gi}, for example) that appear at 1-loop order when
considering the couplings of a scalar to gauge bosons via heavy
fermion or spin-1 loops:
\bea
A_{1/2}(\tau) &=& 2 [\tau + (\tau -1) f(\tau)] \tau^{-2}~, \\ [0.3em]
A_{1}(\tau) &=& - [2 \tau^2 + 3 \tau + 3(2\tau - 1) f(\tau)] \tau^{-2}~,
\eea
where
\bea
f(\tau) &=& 
\left\{
\begin{array}{ll}
{\rm arcsin}^2 \sqrt{\tau}  &  \tau \leq 1   \\ [0.4em]
- \frac{1}{4} \left[ \log \frac{1 + \sqrt{1 - \tau^{-1}}}{1 - \sqrt{1 - \tau^{-1}}} - i \pi \right]^2 &  \tau > 1
\end{array}
\right.
~.
\label{floops}
\eea
In the limit that $\tau \rightarrow 0$, $A_{1/2}(\tau) \rightarrow
4/3$ and $A_{1}(\tau) \rightarrow -7$.

\bibliographystyle{JHEP} 

\bibliography{SO5_biblio}

\end{document}